\newcommand{\unit}[1]{\ensuremath{\, \mathrm{#1}}}

\documentclass[twocol]{ametsoc}

\journal{jas}

\bibpunct{(}{)}{;}{a}{}{,}

\usepackage{color}

\title{\textbf{\large{Why eddy momentum fluxes are concentrated in the upper troposphere}}}

\authors{Farid Ait-Chaalal\correspondingauthor{Farid Ait-Chaalal, Geological Institute, ETH, NOG 60, Sonnegstrasse 5, 8092, Z\"urich, Switzerland}}
\affiliation{ETH Z\"urich, Z\"urich, Switzerland}
\email{farid.chaalal@erdw.ethz.ch}

\extraauthor{Tapio Schneider}
\extraaffil{ETH Z\"urich, Z\"urich, Switzerland, and California Institute of Technology, Pasadena, USA}

\abstract{The extratropical eddy momentum flux (EMF) is controlled by generation, propagation, and dissipation of large-scale eddies and is concentrated in Earth's upper troposphere. An idealized GCM is used to investigate how this EMF structure arises. In simulations in which the poles are heated more strongly than the equator, EMF is concentrated near the surface, demonstrating that surface drag generally is not responsible for the upper-tropospheric EMF concentration.  Although Earth's upper troposphere favors linear wave propagation, quasi-linear simulations in which nonlinear eddy-eddy interactions are suppressed demonstrate that this is likewise not primarily responsible for the upper-tropospheric EMF concentration. The quasi-linear simulations reveal the essential role of nonlinear eddy-eddy interactions in the surf zone in the upper troposphere, where wave activity absorption away from the baroclinic generation regions occurs through the nonlinear generation of small scales. In Earth-like atmospheres, wave activity that is generated in the lower troposphere propagates upward, then turns meridionally, eventually to be absorbed nonlinearly in the upper troposphere. The level at which the wave activity begins to propagate meridionally appears to be set by the typical height reached by baroclinic eddies. This can coincide with the tropopause height but also can lie below it if convection controls the tropopause height. In the latter case, EMF is maximal well below the tropopause. The simulations suggest that EMF is concentrated in Earth's upper troposphere because typical baroclinic eddies reach the tropopause.}

\begin{document}

\maketitle


\section{Introduction}

\begin{figure}[t]
 \centering\includegraphics[width=19pc]{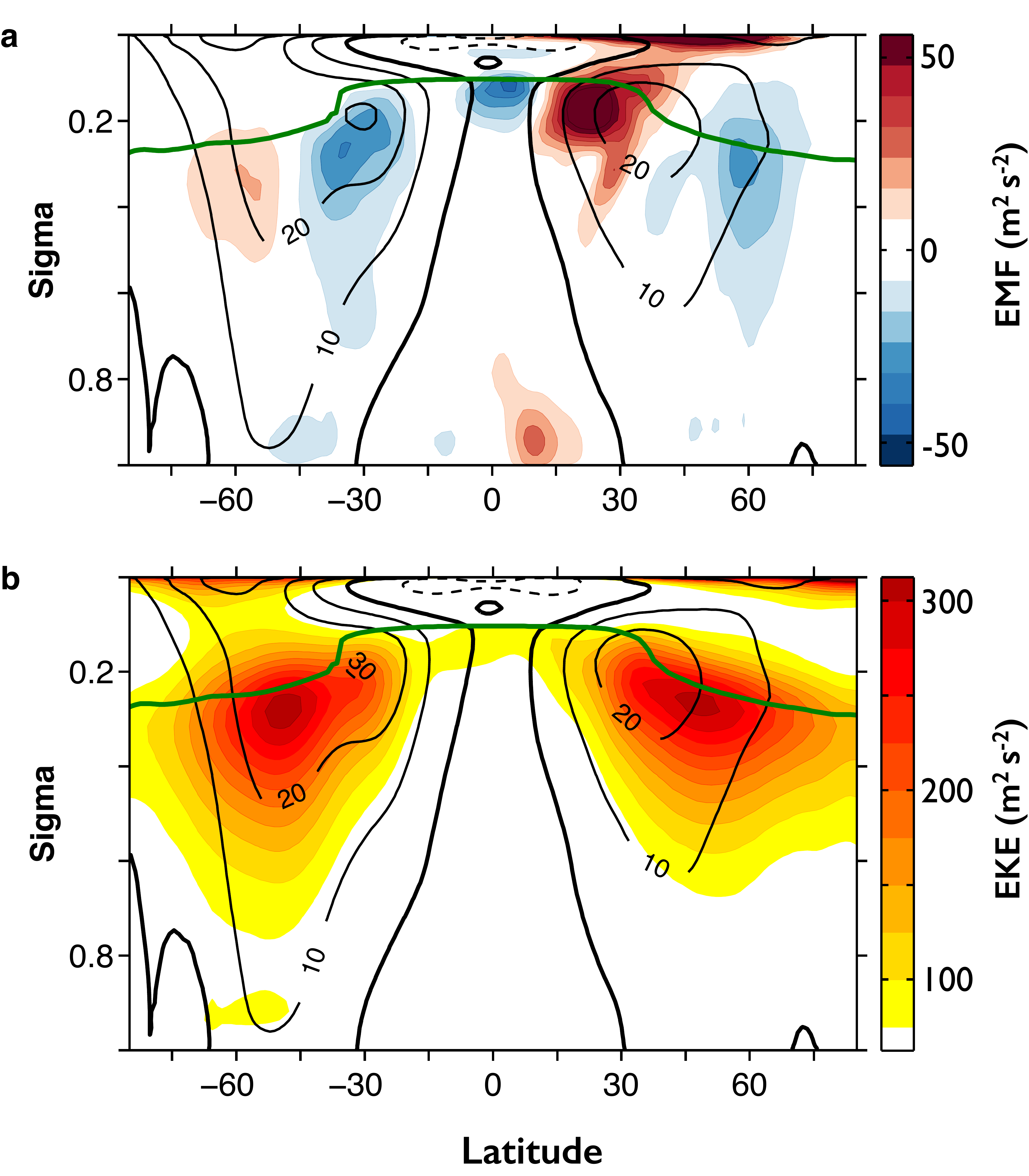}\\
 \caption{Zonal wind (solid lines for eastward and dashed lines for westward winds, in \unit{m\,s^{-1}}) and (a) EMF (colors), (b) EKE (colors). The thick green line is the tropopause (2-\unit{K\,km^{-1}} lapse rate). Based on ERA-40 reanalysis averaged from 1980 to 2001 \citep{uppala2005}.} \label{fig:ERA40}
\end{figure}

Large-scale baroclinic eddies shape the general circulation of Earth's atmosphere. They are generated in midlatitudes through baroclinic instability, propagate meridionally, and dissipate near their critical lines on the flanks of the jet streams \citep{randel1991}. Meridionally propagating eddies transport (angular) momentum toward their generation region \citep{held1975,held2000}. Consequently, generation of large-scale eddies in midlatitudes and dissipation at lower and higher latitudes leads to a meridional momentum flux, with convergence in midlatitudes and divergence in the subtropics and, to a lesser extent, in polar regions (Fig.~\ref{fig:ERA40}a). 

The eddy momentum flux (EMF) controls the structure of the mean zonal surface wind and of meridional cells. To first order in Rossby number, surface friction balances the EMF divergence averaged over an atmospheric column in the extratropics. This balance controls the strength and direction of the mean zonal surface winds. In the upper troposphere, EMF divergence is locally balanced by the Coriolis torque acting on the meridional wind, which accounts for the mass flux in the Ferrel cell, in the high-latitude polar cell, and to some extent in the tropical Hadley cell (see \citealp{schneider2006b} for a review). Hence, the structure of the EMF is fundamental to the mean state of Earth's atmosphere.

It is well known that EMF is concentrated in the upper troposphere, just below the tropopause (Fig.~\ref{fig:ERA40}a). Eddy kinetic energy ($\mathrm{EKE} \propto \overline{u'^2} + \overline{v'^2}$) is also maximal in the upper troposphere (Fig.~\ref{fig:ERA40}b). However, explaining the structures of EMF and EKE is not equivalent because the correlation coefficient between $u'$ and $v'$ varies spatially. The absolute value of the correlation coefficient between  $u'$ and $v'$ generally increases with height within the troposphere, from values smaller than 0.2 above the planetary boundary layer to values around 0.5 in wave-breaking regions near the tropopause. Hence, the concentration of EMF in the upper troposphere is stronger than that of EKE. The structures of EKE and EMF are not the same because EMF arises from the irreversible processes of eddy generation and dissipation, whereas EKE is nonzero even in reversible wave dynamics.
 
The upper-tropospheric concentration of EMF is one of the most conspicuous features of atmospheric eddy fields. Yet proposals of how it arises are scant. It has been proposed that friction acting on the eddies plays a role in reducing eddy amplitudes and meridional propagation near the surface, thus leading to reduced EMF near the surface \citep{held2000,vallis2006}. But observations of Jupiter and Saturn show that friction is unlikely to be important. On Jupiter and Saturn, EMF has been observed by tracking visible clouds in the upper troposphere, with EMF convergence in prograde (westerly) jets and divergence in retrograde (easterly) jets \citep{salyk2006,delgenio2007}. Although we do not have direct observations below the visible clouds, one can infer that EMF must be concentrated in a relatively shallow layer in the upper troposphere, because otherwise the implied transfer of EKE to mean-flow kinetic energy would, implausibly, exceed the total energy available to drive the flow \citep{Schneider2009,liu2010}. Indeed, simulations of Jupiter's and Saturn's atmospheres exhibit EMF concentrated in the upper troposphere, near the tropopause, much like in Earth's atmosphere, although in the simulations this is far above any drag layer, as it likely is for the actual giant planets where drag may arise magnetohydrodynamically in the planet's interior \citep{Schneider2009,liu2010}. So upper-tropospheric concentration of EMF is not unique to Earth's atmosphere but appears to be an ubiquitous feature of planetary atmospheres. And friction is unlikely to be generally responsible for the concentration, as it does not seem to play a role in giant planet atmospheres.

Understanding the EMF concentration requires understanding the generation, propagation, and dissipation of wave activity, as pioneered in studies of baroclinic lifecyles, which revealed the central role of baroclinic growth followed by barotropic decay \citep{simmons1978,simmons1980,heldhoskins1985,thorncroft1991}. Wave activity generation, propagation and dissipation can be diagnosed using cross sections of the Eliassen-Palm (EP) flux \citep{edmon1980}. For dynamics that are locally quasigeostrophic (QG), the meridional and vertical components of the zonal-mean EP flux in pressure coordinates take the form
\begin{equation}\label{E:EP}
\mathbf{F}= R\cos \phi \left(
    \begin{array}{ll}
        -\overline{u'v'} \\
                         \\
        f\,\overline{v'\theta'}/\partial_p \Bar{\theta}
    \end{array}
\right)\mbox{.}
\end{equation}
Here, $R$ is Earth's radius, $\phi$ is latitude, $f$ the Coriolis parameter, $\theta$ potential temperature, and $p$ pressure. Primes denote departures from the zonal average $\overline{(\cdot)}$. The wave activity $A$ obeys 
\begin{equation}\label{E:wave_activity}
       \frac{\partial A}{\partial t} + \nabla \cdot \mathbf{F} = \mathcal{D}\mbox{,}
\end{equation}
where $\mathcal{D}$ includes all non-conservative terms. Equation (\ref{E:wave_activity}) has been shown to hold for small-amplitude waves, in which case the wave activity $A$ equals the pseudomomentum \citep{andrews1976}, and for finite-amplitude waves \citep{nakamura2010}. When WKBJ theory is applicable, $\mathbf{F} \approx \mathbf{c}_g A$ is the advective flux of wave activity that is carried by the group velocity vector $\mathbf{c}_g$ of the waves \citep{lighthill1960,hayes1977}. In general, $\mathbf{F}$ indicates wave activity propagation. In a statistically stationary state, divergence of $\mathbf{F}$ indicates wave activity generation, and convergence of $\mathbf{F}$ wave activity dissipation. 

The meridional component of the EP flux is the meridional eddy angular momentum flux, which is the topic of this paper. In this sense, EMF is linked to meridional wave activity propagation. EMF divergence implies convergence of the meridional wave activity flux, or wave activity dissipation. The EP flux provides a framework for connecting EMF divergence in the upper troposphere to the upward propagation of wave activity generated at lower levels.

It has been suggested that EMF is concentrated in the upper troposphere because potential vorticity (PV) gradients there are greater than in the lower troposphere, which favors meridional wave propagation \citep{held2000,held2007}. The larger PV gradients and stronger zonal winds aloft lead to a wider region in which Rossby wave refractive indices are positive and meridional wave propagation is possible. That is, the critical latitudes, near which meridionally propagating wave activity dissipates  \citep{randel1991}, are farther away from the wave activity generation region \citep{thorncroft1991}. The weaker PV gradients and weaker zonal winds close to the surface, it is suggested, lead to nonlinear saturation of baroclinic eddies close to their generation latitude, precluding substantial EMF across latitudes. But the increase with height of the width of the region that allows meridional wave propagation is more gradual than the peaked structure of EMF: the meridional distance between the critical latitudes widens gradually with height, roughly like the zonal wind contours (Fig.~\ref{fig:ERA40}a). Moreover, the zonal flow on the giant planets is thought to have an approximately barotropic structure in the upper troposphere, so that critical latitudes of waves do not vary strongly with depth. Yet EMF appears to be peaked in the upper troposphere \citep{liu2014}. Such arguments based on linear wave propagation, therefore, also do not seem to account for the entire EMF structure. Even if they would, the question would remain how the mean-flow structures come to be organized in such a way that EMF becomes concentrated in the upper troposphere. The latter question also arises in the context of linear stochastic models that describe fluctuations around a prescribed mean flow \citep{farrell1996,farrell1996b}. Such models are successful in reproducing midlatitude eddy statistics, including an upper-tropospheric EMF enhancement \citep{whitaker1998,zhang1999,delsole2001}. But the mean flow in these models is prescribed rather than influenced by the stochastic eddies, and characteristics of the stochastic forcing are fit, for example, to observations or GCMs. So it likewise is not clear how the upper-tropospheric EMF concentration arises.

Despite their shortcomings, these appear to be the only two hypothesis that have been formulated to explain the upper-tropospheric EMF concentration: surface friction, and the greater linearity of the upper troposphere. In the present paper, we test these two hypotheses explicitly with the help of an idealized dry GCM, which captures the upper-tropospheric EMF concentration and allows us to investigate it systematically. The GCM is described in section~\ref{sec:GCM}. Section~\ref{sec:surf_fric} tests the friction hypothesis by discussing a circulation in which the poles are heated and the tropics are cooled, which exhibits EMF concentrated in the lower troposphere, although surface friction acts there. Section~\ref{sec:noe} tests the linearity hypothesis by comparing fully nonlinear simulations with quasi-linear (QL) simulations, in which nonlinear eddy-eddy interactions are suppressed. We show that nonlinear eddy saturation is more significant in the upper than in the lower troposphere and that eddy-eddy interactions are essential for capturing eddy absorption. In section~\ref{sec:mech}, we propose a mechanism that accounts for the upper-tropospheric EMF concentration, focusing on the typical depth of baroclinic eddies, on the tropopause as a wave guide, and on nonlinear wave saturation in the upper troposphere. 

\section{Idealized GCM}\label{sec:GCM}

The idealized dry GCM is based on the pseudospectral dynamical core of the Geophysical Fluid Dynamics Laboratory's Flexible Modeling System. It is described in detail in \cite{schneider2006}. The only  modification we adopt is the upgrade from the Robert-Asselin filter to the Robert-Asselin-Williams filter in the leapfrog time stepping \citep{williams2010}. The primitive equations on a sphere are integrated using the pseudospectral method on unevenly spaced vertical $\sigma$-levels \citep{bourke1974,simmons1981}. 

The model is thermally driven by Newtonian relaxation of temperatures toward the radiative equilibrium of a semi-gray atmosphere. The surface temperature $T_s^e$ in radiative equilibrium is given as a function of latitude $\phi$ by
\begin{equation}\label{E:forcing}
	T_s^e(\phi)=\widetilde{T}_s^e + \Delta_h \cos^2 \phi \mbox{,}
\end{equation}
where $\Delta_h$ is the pole-to-equator temperature contrast and $\widetilde{T}_s^e$ the radiative-equilibrium surface temperature at the poles. For Earth-like simulations, we set $\Delta_h=90$ K and $\widetilde{T}_s^e=260\, \unit{K}$, with $\overline{T}_s^e=\widetilde{T}_t^e + 2\Delta_h/3=320\, \mathrm{K}$ the corresponding global-mean surface temperature in radiative equilibrium. The radiative-equilibrium skin temperature at the top of the atmosphere is set to $T_t^e=200\, \unit{K}$. The radiative-equilibrium temperature $T^e$ decreases monotonically with altitude and is given as function of the optical depth $d_0(\phi)$ and 
the pressure $p$ by
\begin{equation}
	T^e(p,\phi)=T_s^t\left[1 + d_0(\phi)\left(\frac{p}{p_0}\right)^{\alpha}\right]^\frac{1}{4} \mbox{,}
\end{equation}
where $\alpha=3.5$ is approximately the ratio of absorber (e.g., water vapor) scale height to density scale height. The optical depth is specified as
\begin{equation}
d_0(\phi)=\left[\frac{T_s^e(\phi)}{T_t^e}\right]^4-1,
\end{equation}
to achieve a continuous monotonic decrease of radiative-equilibrium temperatures from the surface to a constant temperature at the top of the atmosphere. The Newtonian relaxation time scale $\tau$ is a function of latitude $\phi$ and vertical coordinate $\sigma$, as in \cite{held1994},
\begin{equation}
	\tau(p,\phi)=\tau_i^{-1}+(\tau_s^{-1}-\tau_i^{-1})\max\left(0,\frac{\sigma-\sigma_b}{1-\sigma_b} \right)\cos^8(\phi).
\end{equation}
The constants $\tau_i$ and $\tau_s$ are the relaxation times in the interior of the atmosphere and at the surface in low latitudes. Earth-like simulations are carried out with $\sigma_b = 0.85$, $\tau_i = 50$~days and $\tau_s = 10$~days. The Newtonian relaxation scheme is presented in more detail in \cite{schneider2004b}.

The radiative-equilibrium temperature profile $T_s$ is statically unstable in the lower atmosphere. A convective parametrization redistributes enthalpy vertically and mimics the stabilizing effect of latent heat release in moist convection. When an atmospheric column is statically less stable than a prescribed lapse rate $\gamma \Gamma_d$, with dry adiabatic lapse rate $\Gamma_d = g/c_p$ and $\gamma \le 1$, its temperature is relaxed toward $\gamma \Gamma_d$ on a time scale of 4~hours. Details can be found in \cite{schneider2006}. The implied (but not explicit) latent heat release increases as $\gamma$ decreases; a value $\gamma = 1$ corresponds to vertical entropy homogenization through dry convection. Earth-like simulations are performed with $\gamma=0.7$, corresponding to a lapse rate of $6.9 \unit{K\,km^{-1}}$.

Dissipation consists of  $\nabla^8$ hyperviscosity acting on temperature, vorticity, and divergence, and of momentum and dry static energy diffusion in a 2.5-km deep boundary layer \citep{smagorinsky1965}.

The simulations are performed at horizontal spectral resolution T85 with 30 vertical $\sigma$-levels. All time averages are performed over $600$ days after 1400 days of spin-up.

\section{Heating the poles and cooling the tropics}\label{sec:surf_fric}

Latitude dependent radiative forcing on Earth introduces a vertical asymmetry of the troposphere because, at leading order, zonal-mean vertical shear is proportional to zonal-mean meridional temperature gradients (thermal wind blance). Here we examine a circulation in which the pole-to-equator surface temperature gradient is reversed. This is achieved by setting $\Delta_h = -90\, \unit{K}$ and $\widetilde{T}_s^e = 380\, \unit{K}$; that is, radiative-equilibrium temperatures near the poles are larger than near the equator. The Newtonian relaxation time scale to this radiative-equilibrium temperature is uniform with $\tau_s = \tau_i = 40$ days. All other GCM parameters are unchanged from the Earth-like simulation introduced in section~\ref{sec:GCM}. 

\begin{figure*}[ht!]
 \centering\includegraphics[width=\textwidth]{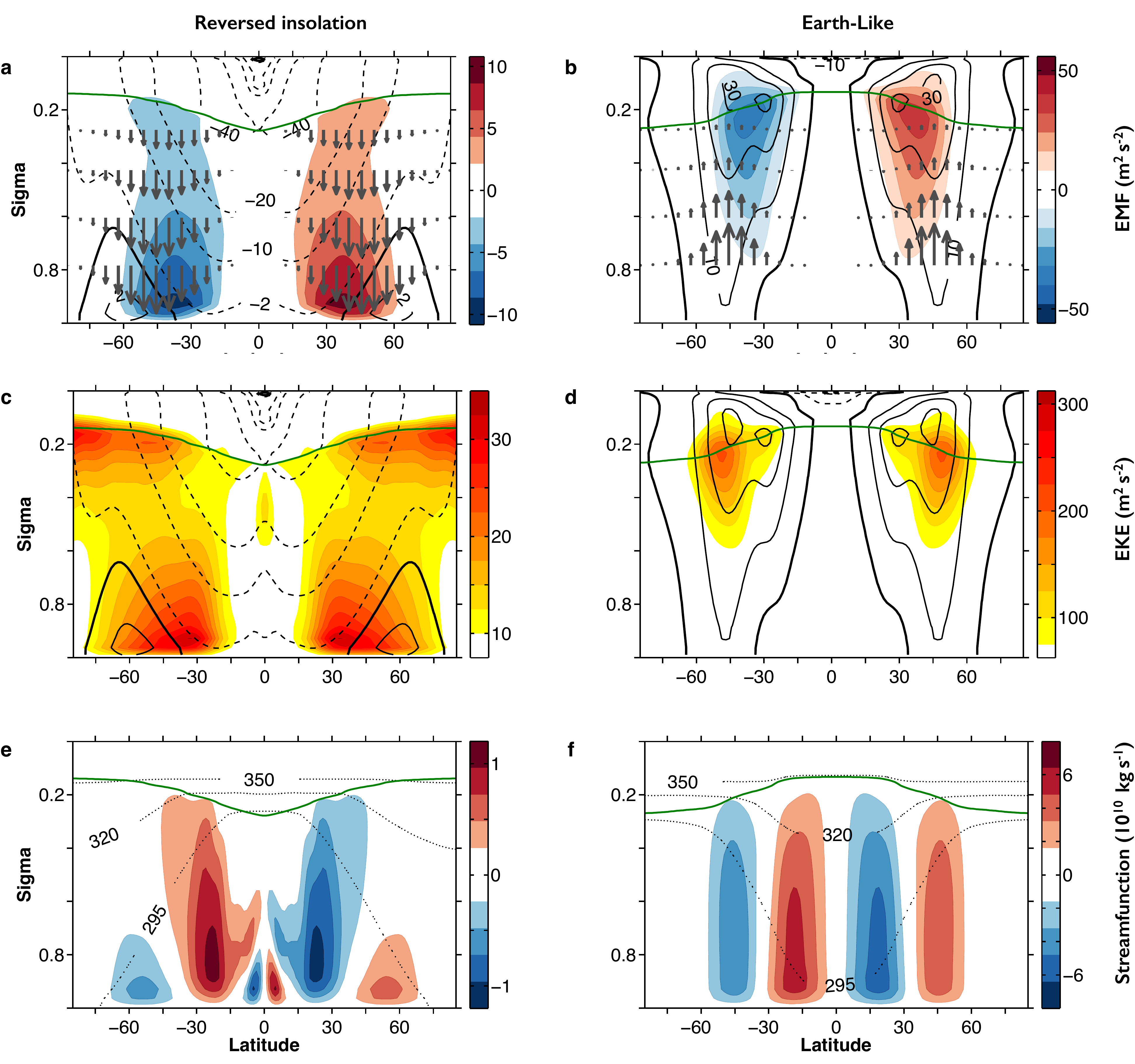}\\
 \caption{Comparison of a circulation in which the poles are heated and the tropics cooled (left column) with an Earth-like simulation (right column). Colors show (a, b) EMF, (c, d) EKE, and (e, f) the mass streamfunction. The solid and dashed lines in (a-d) indicate westward and eastward zonal wind in $\unit{m\,s^{-1}}$, and the dark grey arrows the vertical component of the EP flux. Note that color scales are different in the reversed insolation and in the Earth-like configurations. The dotted lines in (e, f) represent three isentropes (295\,\unit{K}, 320\,\unit{K}, and 350\,\unit{K}) as an indicator of the thermal structure. The thick green line marks the tropopause (a 2-\unit{K\,km^{-1}} lapse rate contour). The statistically identical northern and southern hemispheres are averaged, in addition to averaging zonally and temporally (over 600 days after a 1400 days spin-up to reach a statistically steady state).}\label{fig:inv_delh}
\end{figure*}

The EMF structure (Fig.~\ref{fig:inv_delh}a) is an upside-down version of the Earth-like structure (Fig.~\ref{fig:inv_delh}b). Maximal EMF occurs inside the atmospheric boundary layer, where drag is acting on the flow. Surface drag is known to affect properties of macroturbulent eddies and the general circulation, including EMF amplitude, jet strength, and jet location \citep{james1987,robinson1997,chen2007,liu2014}. It has also been suggested that surface drag might at least partially explain the upper-tropospheric concentration of EMF \citep{held2000,vallis2006}. Our GCM simulation indicates that friction alone is not responsible for the upper-tropospheric concentration of EMF, because the maximal convergence occurs in the frictional boundary layer when temperature gradients are reversed. 

Baroclinic eddies in the simulation are generated, as on Earth, in the extratropical troposphere, as can be diagnosed from the time evolution of a simulation started from a slightly perturbed axisymmetric state with zonal winds and temperature structure equal to the zonal mean of the statistically steady state shown above. The Charney-Stern (\citeyear{charney1962}) necessary condition for baroclinic instability is satisfied: although surface potential temperatures increase poleward, so that the surface temperature gradient is reversed relative to Earth's, the interior-tropospheric PV gradient is also reversed (negative) in midlatitudes. The quasigeostrophic potential vorticity (QGPV) gradient along isobars approximates the PV gradient along isentropes \citep{charney1962} and is given by 
\begin{equation}\label{E:grad_QGPV}
 \partial_y \bar{q} = \beta - \partial_{yy} \bar u + f \, \partial_p \left( \frac{\partial_y \Bar\theta}{\partial_p \Bar\theta} \right).
\end{equation}
The reversal of the QGPV gradient in the interior troposphere arises because the stretching term (third term on the right-hand side) dominates the planetary vorticity gradient $\beta = \partial_y f$, where $y=R \phi$ is the meridional distance coordinate. In other words, the slope of the isentropes, $I = - \partial_y \Bar\theta/\partial_p \Bar\theta$, flattens sufficiently rapidly with altitude in the interior troposphere that the stretching term reverses the sign of the QGPV gradient. In polar regions and in the subtropics, the QGPV gradient has the same sign as $\beta$, so that the Charney-Stern criterion for baroclinic instability is satisfied only in midlatitudes.

For baroclinic instability in a QG flow, the divergence of the zonal momentum flux is equal to a temporally growing exponential times a weighted vertical integral of the QGPV meridional gradient, with a strictly positive weight \citep{held1975}. As a consequence, linear theory points to EMF divergence where the QGPV gradient is positive throughout the column, that is, in our case, outside the baroclinic zone in midlatitudes. Angular momentum conservation then implies EMF convergence within the baroclinic zone. 

Indeed, in our simulation with reversed temperature gradient, like on Earth, EMF is converging in midlatitudes and diverging at low-latitudes (Fig.~\ref{fig:inv_delh}a); that is, EMF is directed poleward between the subtropical dissipation regions and the baroclinic generation region. Because of surface drag, EMF convergence in midlatitudes results in westerly surface winds, and divergence results in easterly surface winds closer to the equator. Westerlies and easterlies are associated  with the analogs of Earth's Ferrel and Hadley cells, as indicated by the Eulerian streamfunction in Fig.~\ref{fig:inv_delh}e (to be comapred with the Earth-like situation in Fig.~\ref{fig:inv_delh}f). Ferrel cells are here thermally direct but eddy driven and very shallow because EMF is concentrated close to the surface. Hadley cells are thermally indirect and also eddy driven. Close to the surface, two thermally direct and thermally driven cells can be seen equatorward of $15^{\circ}$. 

Reversed meridional temperature gradients imply that eddy entropy fluxes, which are generally directed down the potential temperature gradient (e.g., \citealp{kushner1998,held1999b}), are in the mean equatorward. Because the vertical component of the EP flux (\ref{E:EP}) is proportional to the eddy potential temperature flux, it is directed downward. This is illustrated in Figs.~\ref{fig:inv_delh}a, b. It implies downward, rather than the usual upward, propagation of wave activity \citep{edmon1980}. Wave activity then accumulates near the surface in the extratropics, propagates equatorward and dissipates, leading to EMF and EKE maxima there, despite the drag (Fig.~\ref{fig:inv_delh}a,c). Wave activity appears to propagate horizontally where farther vertical propagation is inhibited, be it by a solid boundary in the case of reversed temperature gradients or by the tropopause, an interface at which the static stability increases, in the Earth-like case. 

This simulation shows that upper-level EMF enhancement for an Earth-like simulation and the underlying asymmetry between the  lower and the upper part of the troposphere cannot be attributed to surface friction alone. We have verified that this conclusion is robust when the strength of surface friction is varied. The latter has a strong effect on the near-surface zonal and eddy kinetic energies, but not on the EMF strength. Similarly, we have verified that the conclusions continue to hold when surface friction is only applied to the zonal-­mean flow but not to the eddies. Both in Earth-like and reversed insolation simulations, the EMF structure is qualitatively unchanged when surface friction acts only on the mean flow, consistent with previous studies of the separate effects of friction on mean flow and eddies \citep{chen2007}. In both cases, we observe that EMF strength is approximately doubled when surface friction acts only on the mean flow. This likewise shows that frictional damping of eddies in Earth­-like situations is not responsible for the upper-­tropospheric EMF enhancement.

Our reversed insolation experiment was motivated by theoretical considerations. Nevertheless, it is worthwhile noting that it can describe the atmosphere of planets with high obliquity, which were recently studied by \cite{ferreira2014}.

\section{Quasi-linear simulations}\label{sec:noe}

As explained earlier, it has been suggested that an upper troposphere that is more linear than the lower troposphere favors meridional propagation of eddies and may explain upper-tropospheric EMF enhancement \citep{held2007}. To test this hypothesis, we compare fully nonlinear to QL simulations for Earth-like parameters.  A QL model only captures the linear behavior of eddies and their nonlinear interaction with a mean flow. Nonlinear eddy-eddy interactions are suppressed. 

\subsection{Quasi-linear model}\label{sec:ee_removal}

Eddies are defined as departures from the average over longitude $\lambda$ at constant latitude $\phi$ (zonal mean): 
\begin{equation}
	a(\phi,\lambda,p)=\bar{a}(\phi,p) + a'(\phi,\lambda,p) \mbox{.}
\end{equation}
Throughout this paper, the overbar denotes a zonal average at constant pressure $p$ and the prime the departure from this average. In the GCM, we use a surface pressure-weighted zonal average ${\overline{p_s (\cdot)}}/{\bar{p_s}}$ along $\sigma$ surfaces because surface pressure acts as a density in $\sigma$ coordinates (whereas the flow is non-divergent in pressure coordinates). 

The QL approximation means keeping eddy-mean flow interactions and removing eddy-eddy interactions. The removal of eddy-eddy interactions in the dry GCM consists of modifying the momentum and thermodynamic equations as described by \cite{ogorman2007}. For example, the time tendency owing to meridional advection of a field $a=\bar{a}  + a'$ by a velocity $v=\bar{v}  + v'$ can be decomposed as 
\begin{align}\label{E:advection}
	\frac{\partial a}{\partial t} &=-v\frac{\partial a}{\partial y} \notag \\ 
     &=-\bar{v}\frac{\partial \bar{a}}{\partial y} -
	\bar{v}\frac{\partial a'}{\partial y}-
	v'\frac{\partial \bar{a}}{\partial y}-
	v'\frac{\partial a'}{\partial y} \mbox{.}
\end{align}
The term $v'{\partial a'}/{\partial y}$ in (\ref{E:advection}) represents the advection of the eddies by the eddies themselves.  It can be decomposed into a mean part $\overline{v'{\partial a'}/{\partial y}}$ and a fluctuating part. The removal of the eddy-eddy interactions consists of keeping only the former in (\ref{E:advection}):
\begin{align}\label{E:advection_noe}
	\frac{\partial a}{\partial t} &=-v\frac{\partial a}{\partial y} \notag \\ 
     &\stackrel{\mathrm{QL}}{\approx} -\bar{v}\frac{\partial \bar{a}}{\partial y}-
	\overline{v}\frac{\partial a'}{\partial y}- 
	v'\frac{\partial \bar{a}}{\partial y}-
	\overline{v'\frac{\partial a'}{\partial y}}\mbox{.}
\end{align}
As a consequence, all triad interactions involving only eddy quantities are removed, such that interactions between the eddies and the mean flow are the only nonlinear interactions retained. This approach has received some attention since early studies of rotating large-scale flows. For example, small-amplitude wave activity conservation theorems have been derived within its scope \citep{charney1961, eliassen1961,dickinson1969,boyd1976,andrews1976,andrews1978}. Keeping only eddy--mean flow interactions in the tendency equation (\ref{E:advection_noe}) is equivalent to linearizing the equation for the fluctuating part $a'$, keeping the equation for $\bar{a}$ unchanged. Hence the ``quasi-linear'' denomination for the resulting set of equations. However, it does not imply that all nonlinearities are small because nonlinear eddy--mean flow interactions can be strong.

The QL approximation is well-posed in the sense that it preserves inviscid invariants consistent with the order of the truncation: mass, angular momentum, entropy, energy, entropy variance and, when applicable, potential enstrophy. First-order invariants are conserved because first moment equations are unchanged, and second-order invariants are conserved because neglected third-order terms only redistribute second-order inviscid invariants among scales. 

In terms of a statistical closure, the moment or cumulant equations implied by the QL equations are closed at second order; third-order cumulants do not enter the second-order equations. So the QL approximation corresponds to truncating the hierarchy of cumulant equations at second order---an approximation that has recently been called CE2 and has been used to study the dynamics of barotropic jets \citep{marston2008,tobias2011,srinivasan2012,marston2014}. Statistical Structural Stability Theory \citep{farrell1996,farrell1996b} and some kinetic theories of statistical physics \citep{bouchet2013} are essentially equivalent to CE2.

\subsection{Mean zonal and meridional circulations}\label{sec:mmc}

\citet{ogorman2007} described the mean zonal circulation in the QL model and compared it with that in the corresponding fully nonlinear simulation. We extend here some of their findings to the EMF structure and the mean meridional circulation.

\begin{figure}[t]
 \centering\includegraphics[width=19pc]{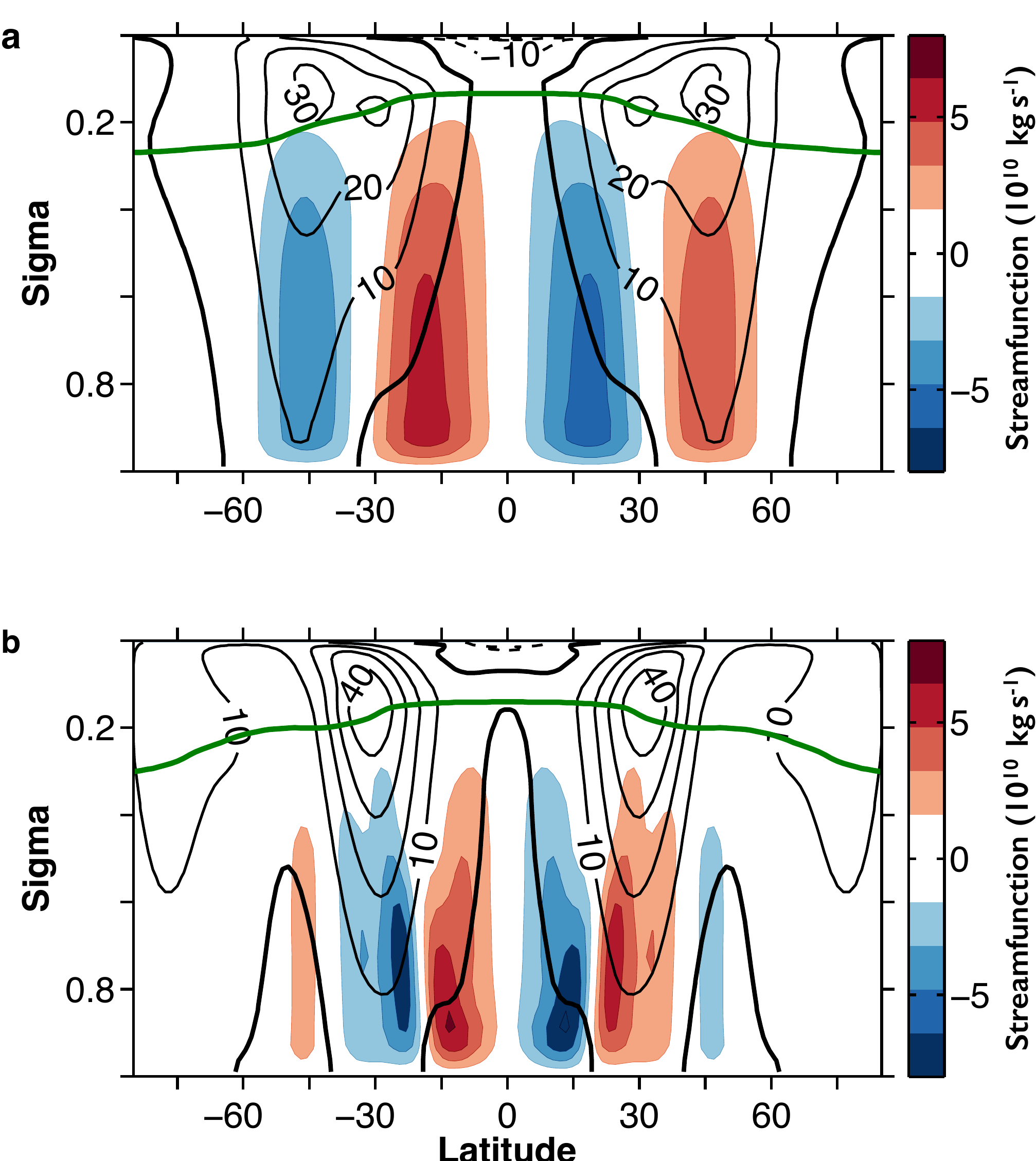}\\
 \caption{Mass flux streamfunction (colors) and zonal wind in \unit{m\,s^{-1}}. (a) Full model. (b) QL  model. The thick green line marks the tropopause. As in Fig.~\ref{fig:inv_delh}, the fields are averaged zonally, temporally and over both hemispheres.} \label{fig:stream_wind}	
\end{figure}

Figure~\ref{fig:stream_wind} shows the mean zonal wind and Eulerian mean meridional streamfunction. The QL model produces upper-atmospheric jet streams above surface westerlies ("eddy-driven jets"), associated with meridional cells extending from the surface to the tropopause. The circulation in the QL approximation is compressed in the meridional direction compared with the full model.  The mid-latitude jet has a limited meridional extent and a secondary jet appears at higher latitudes. This is consistent with the Eulerian streamfunction, which exhibits narrower Hadley and Ferrel cells. Moreover, weaker cells appear at higher latitudes, in association with the secondary jets, such that four cells occur in each hemisphere (the contouring in Fig.~\ref{fig:stream_wind} does not reveal the weak high-latitude cells). \cite{ogorman2007} attributed the meridional compression of the meridional circulation to the fact that the suppressed eddy-eddy interactions isotropize the eddies and thus generally increase meridional scales \citep{stone1972,shepherd1987}.

Thus, the QL model successfully reproduces some features of the fully nonlinear general circulation, such as eddy-driven jets and meridional cells, together with a realistic thermal structure (three isentropes are indicated in Fig.~\ref{fig:edmf_wind}). However, significant qualitative and quantitative differences suggest that the redistribution of angular momentum by large-scale eddies is not realistic. We now examine the structure of the EMF. 
\subsection{Eddy momentum flux}\label{sec:edmf}
\begin{figure}[t]
  \centering\includegraphics[width=19pc]{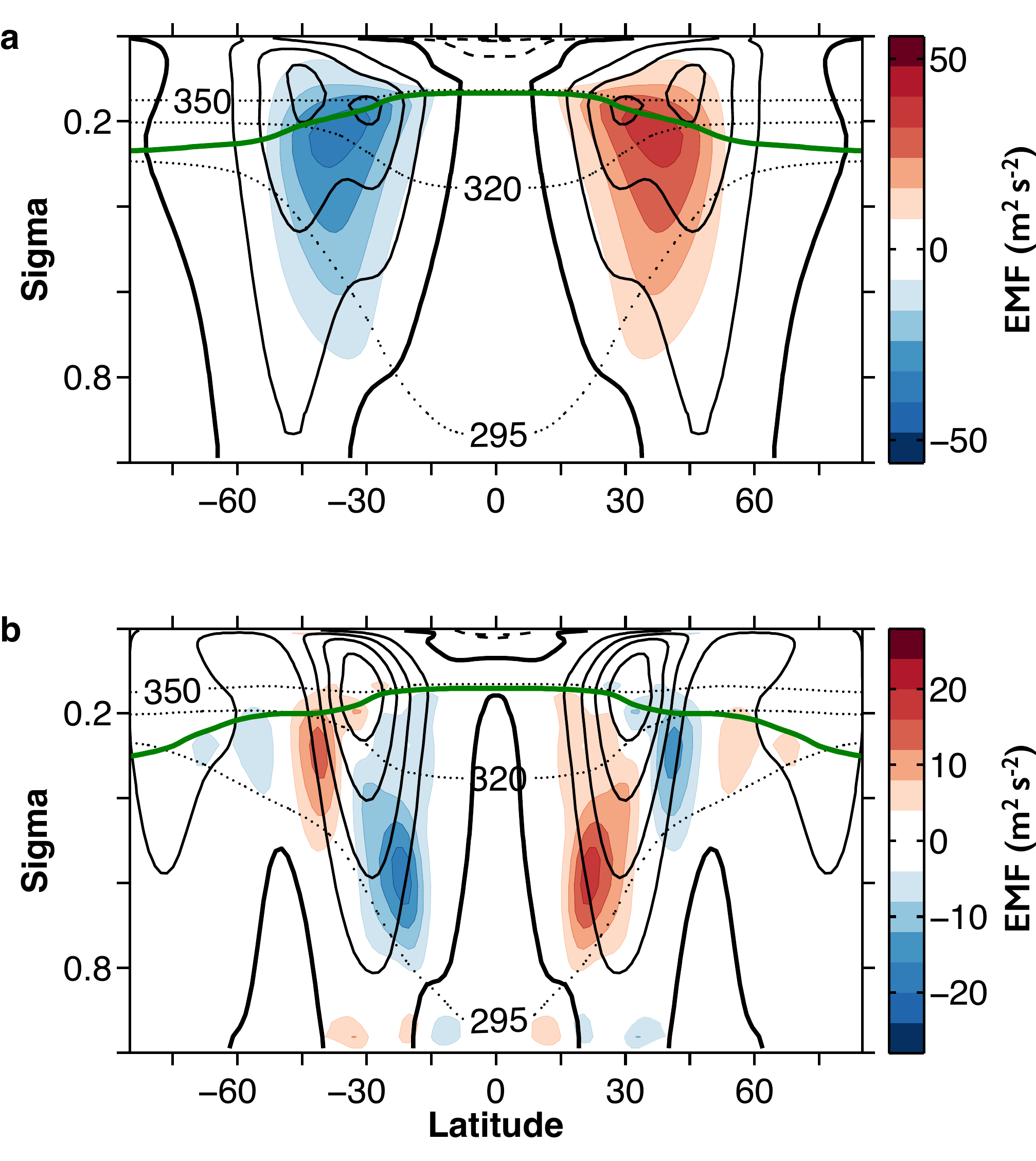}\\
  \caption{EMF (colors) and zonal wind (contours).
          (a) Full model. (b) QL model. Note the different color scales. The EMF amplitude in the full model is about twice as large as in the QL model. The dotted lines represent three isentropes (295 \unit{K}, 320\unit{K}, and 350\unit{K}). The thick green line marks the tropopause (a 2-\unit{K\,km^{-1}} lapse rate contour). As in Fig.~\ref{fig:inv_delh}, the fields are averaged zonally, temporally and over both hemispheres.}\label{fig:edmf_wind}
\end{figure}
Averages of EMF in the full model and in the QL model are compared in Fig.~\ref{fig:edmf_wind}. The top panel (Fig.~\ref{fig:edmf_wind}a) depicts the well-known picture of momentum transported by eddies in the upper troposphere from the subtropics into midlatitudes. Hence, EMF is converging above midlatitude surface westerlies and diverging above low-latitude surface easterlies. Idealized dry dynamics with a uniform surface reproduce the essence of the zonal flow and EMF structure in Earth's troposphere (Fig.~\ref{fig:ERA40}a). 

The bottom panel (Fig.~\ref{fig:edmf_wind}b) shows that in the QL model, EMF exhibits a fundamentally different pattern. Consistent with surface westerlies around $30^\circ$ and $75^\circ$, EMF is converging in the atmospheric column above. However, instead of a poleward EMF over much of the baroclinic zone like in the full simulation, we observe in the QL simulation equatorward and poleward transport on the poleward and equatorward flanks of the jet respectively. There are no well-defined upper-tropospheric extrema. Poleward EMF is maximal in the mid-troposphere, in sharp contrast to the full simulation.

\begin{figure}[t]
  \centering\includegraphics[width=19pc]{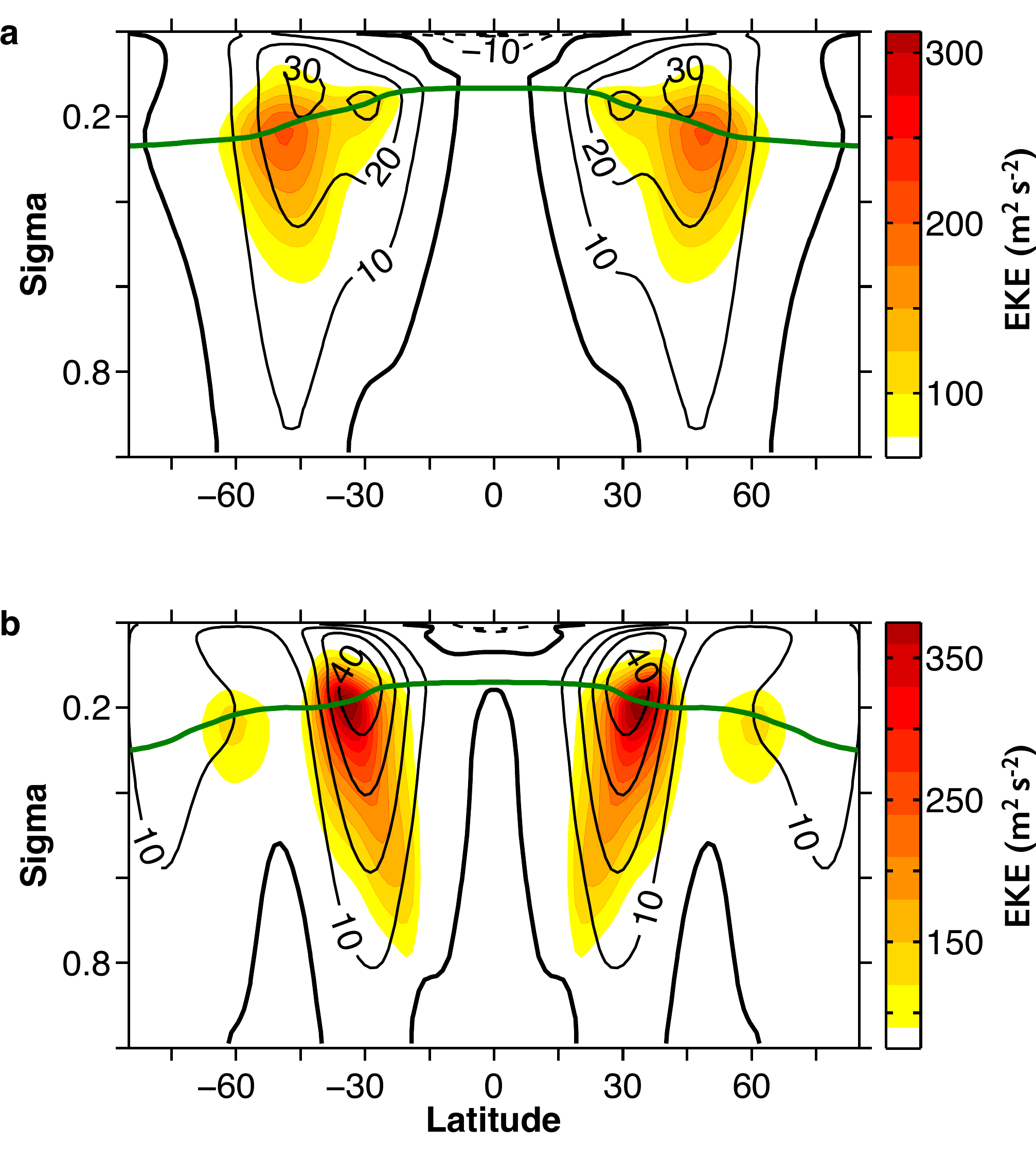}\\
  \caption{EKE (colors) and zonal wind (contours). (a) Full model. (b) QL model. Note the different color scales. The EKE amplitude in the QL model is about twice as large as in the full model. See Fig.~\ref{fig:edmf_wind} for other details. }\label{fig:EKE_wind}
\end{figure}

EKE, shown in Fig.~\ref{fig:EKE_wind}, is maximal in the core of the jet, just below the tropopause, both in the QL and in the fully nonlinear model. Despite weak upper-tropospheric EMF, especially on the equatorward flank of the jet, EKE is still maximal near the tropopause in the QL simulation. Because the ratio of EMF to EKE can be interpreted as a correlation coefficient between meridional and zonal eddy velocity components, this implies that, contrary to the full simulation, $u'$ and $v'$ decorrelate in the QL upper troposphere, especially equatorward of the main mid-latitude jet. Despite being of large amplitude, baroclinic eddies do not transport much angular momentum from the subtropics to midlatitudes. This suggests that the shortcomings of the QL approximation in reproducing the EMF structure do not arise because eddies would not reach the upper troposphere, for example, but because vertical propagation might not be captured adequately. Instead, the QL approximation does not capture the dissipation of eddies.
\begin{figure}[t]
  \centering\includegraphics[width=19pc]{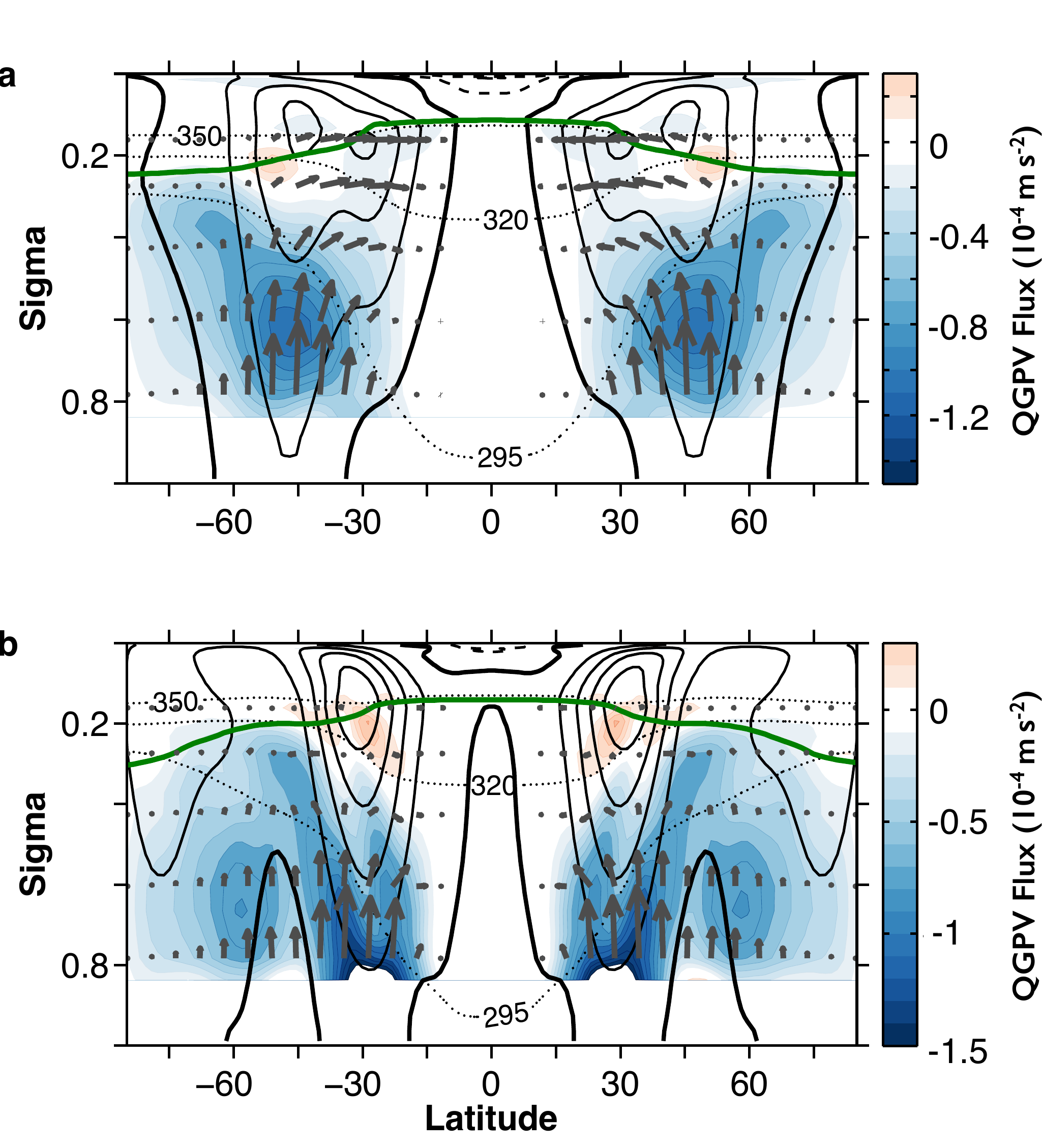}
  \caption{Quasigeostrophic EP vector (grey arrows) and QGPV flux (colors). (a) Full model. (b) QL model. The solid and dashed contours indicate zonal wind (as in Fig.~\ref{fig:edmf_wind}). See Fig.~\ref{fig:edmf_wind} for other details. }\label{fig:vq_EP_wind}
\end{figure}
To obtain a more precise picture of wave activity propagation and dissipation, we compute cross sections of the QG EP vector $\mathbf{F}$ (Eq.~\ref{E:EP}) and of the flux of QGPV $q$, which is proportional to $\nabla \cdot \mathbf{F}$ through the Taylor identity \citep{edmon1980},
\begin{equation}\label{E:gen_taylor}
	\overline{v'q'}=\frac{1}{R \cos\phi} \, \nabla \cdot \mathbf{F}\mbox{.}
\end{equation}
Both $\mathbf{F}$ and $\overline{v'q'}$ are shown in Fig.~\ref{fig:vq_EP_wind}. The meridional extent of lower-tropospheric negative QGPV flux is comparable in the full and the QL models. The EP fluxes are qualitatively similar, roughly below the 295-K isentrope, where the QGPV flux is dominated by the vertical gradient of the meridional eddy flux of potential temperature. Baroclinic growth and vertical propagation of wave activity seem fairly well captured by QL dynamics in this part of the atmosphere.

Betweeen the 320-K isentrope and the tropopause, the EP flux is very weak in the QL model. QGPV fluxes are mostly positive, whereas in the full model, significant meridional EP flux occurs, and QGPV fluxes are negative, indicating absorption of eddies on the equatorward flank of the jet. The lack of absorption in the QL model, and even weak emission as suggested by the positive QGPV flux, accounts for the absence of enhanced EMF in the subtropical upper troposphere. It is consistent with large values of EKE being associated with weak EMF. 

Interestingly, the QL model performs better in the lower troposphere than in the upper troposphere. This appears to contradict the view of a more linear upper troposphere and of a more turbulent lower troposphere \citep{held2007}. However, considering time averages gives little information about the dynamical processes involved. To shed some light on the dynamics of the QL approximation and on the role of large-scale eddy-eddy interactions in eddy absorption, we perform baroclinic wave lifecycle experiments. 

\subsection{Lifecycle experiments}\label{sec:LC}

A lifecycle experiment solves an initial value problem and aims at understanding the development and saturation of a growing disturbance in a baroclinically unstable zonal flow \citep{simmons1978,simmons1980,thorncroft1991,tim2009}. The initial condition we use here is a small-amplitude disturbance of normal-mode form with respect to the mean circulation of the full model (Fig.~\ref{fig:stream_wind}). We choose the zonal wavenumber $k_i=6$ of the disturbance, corresponding to the fastest growing mode for the full model. Radiative and boundary layer parametrizations are disabled. Lifecycle experiments are run for the full and the QL models. In both cases, we use a normal mode of the statistically steady circulation of the fully nonlinear model.  

The initial disturbance is obtained with a version of the GCM linearized around the mean circulation of the full model. We follow a similar procedure as in \cite{tim2009}. As an initial condition for this linear analysis, we perturb at all vertical levels the odd meridional spectral coefficients (from 3 to 83) of the vorticity field, corresponding to the zonal wavenumber $k_i$. Only the spectral modes of all fields corresponding to $k_i$ are advanced in time. Surface pressure is uniformly rescaled such that the amplitude $\Delta p \equiv \langle (p_s - \bar{p_s})^{2} \rangle^{1/2}$ is reset to 1\unit{Pa} when  $\Delta p$ exeeds 10\unit{Pa} (brackets $\langle \cdot \rangle$ indicate a global average and $\overline{(\cdot)}$ a zonal average). Temperature and velocity fields are rescaled accordingly using geostrophic balance. The rescaling procedure is repeated a few times until the disturbance growth rate and phase speed remain constant. 

\paragraph*{Energy cycle}\label{sec:energy_cycle}

Before comparing wave activity propagation in the QL and fully nonlinear lifecycles, we briefly discuss the time evolution of EKE and of the two energy conversions involving EKE: the baroclinic conversion $C_E$ from eddy available potential energy (EAPE) to EKE, and the barotropic conversion $C_K$ from zonal kinetic energy (ZKE) to EKE \citep{lorenz1955}. The baroclinic conversion $C_E$ is significant during baroclinic growth; the barotropic conversion $C_K$ is significant during barotropic decay, corresponding to zonal flow acceleration through decaying eddies \citep{simmons1978}. The dominant term in $C_K$ involves EMF convergence \citep{lorenz1955}.

\begin{figure}[!ht]
 \centering\includegraphics[width=19pc]{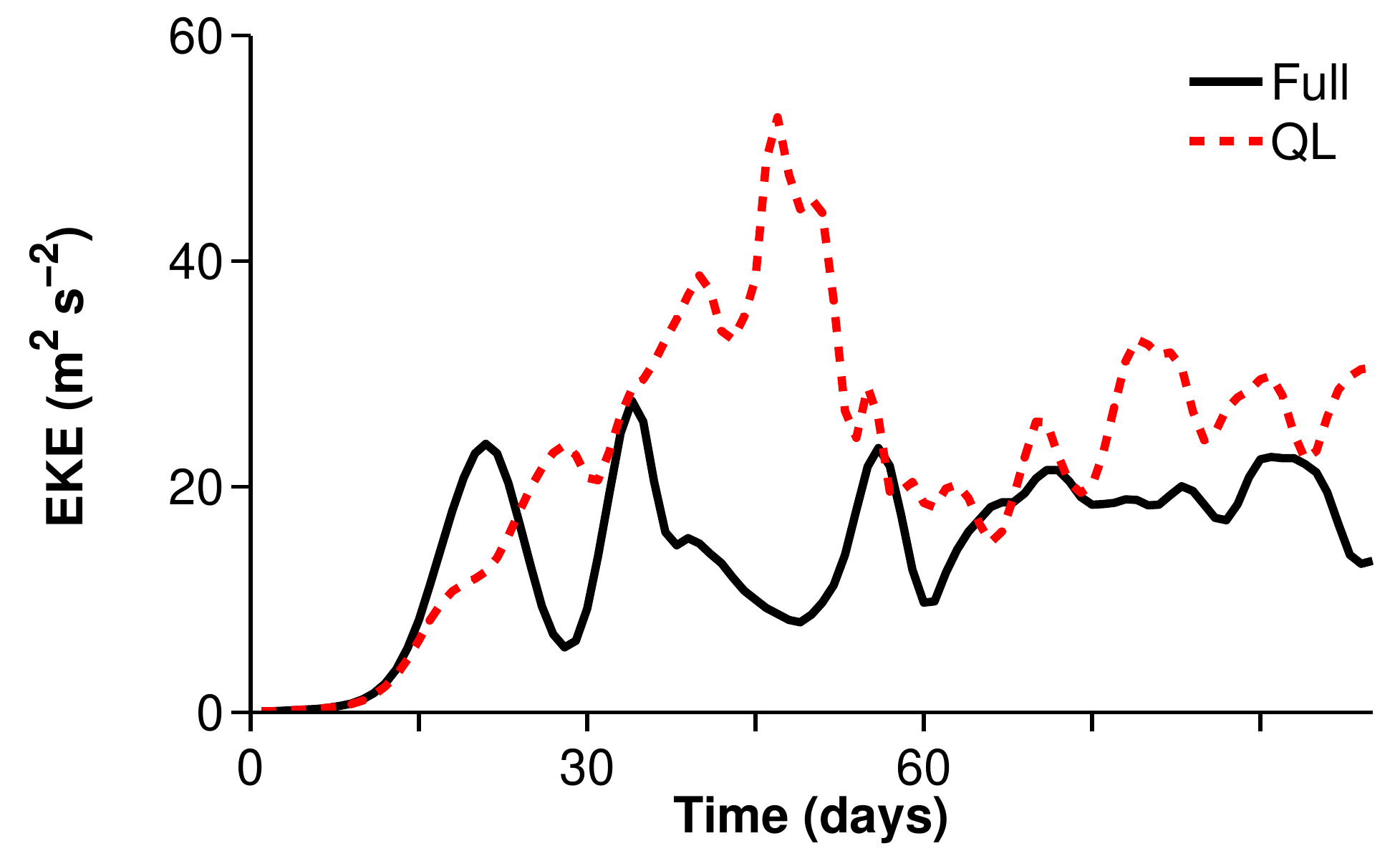}\\
 \caption{Global mean EKE as a function of time in the lifecycle  experiments for the full model (solid black line) and the QL model (dashed red line). }\label{fig:eke}
\end{figure}

\begin{figure}[!ht]
 \centering\includegraphics[width=19pc]{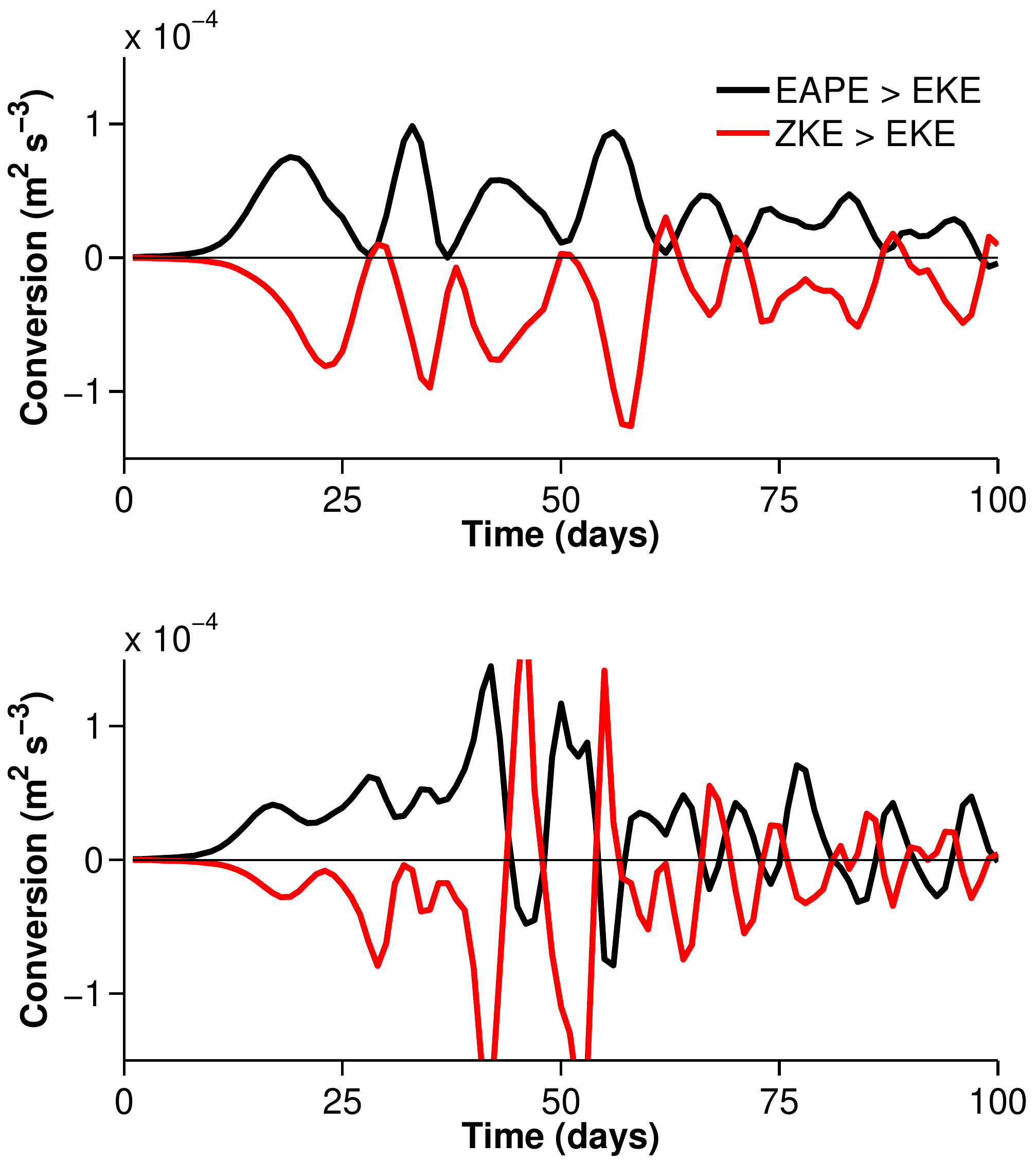}\\
 \caption{Baroclinic (black) and barotropic (red) energy conversions $C_E$ and $C_K$ as a function of  time in the lifecycle experiments for (a) the full model and (b) the QL model.} \label{fig:conv}
\end{figure}

The evolution of EKE as a function of time is shown in Fig.~\ref{fig:eke}; $C_E$ and $C_K$ are shown in Fig.~\ref{fig:conv}. The fully nonlinear simulation exhibits several cycles of growth, saturation, and decay of baroclinic eddies. The two first cycles, from day 0 to day 28 and from day 28 to day 38, correspond to what is discussed in \cite{simmons1978}. However, the QL model does not exhibit as clearly defined lifecycles (Figs.~\ref{fig:eke} and \ref{fig:conv}b); the time evolutions of both $C_E$ and $C_K$ are different. First, there is cyclical large-amplitude conversion from ZKE to EKE, especially after day 45. (Small-amplitude conversion also happens in the full model, for example at day 28; see Fig.~\ref{fig:conv}a.) Second, when $C_K$ is negative, there is also relatively large conversion from EKE to EAPE, which does not occur in the full model.

We now look at wave activity propagation and absorption during barotropic decay and at the termination of lifecycles, to understand why QL lifecycles are so different and to make the connection with the different EMF structures.   

\paragraph*{Wave activity diagnostics}\label{sec:EP_Cross}

We compute cross-sections of the EP flux and of the QGPV flux, as in the lifecycle studies of \cite{thorncroft1991}. For small-amplitude conservative eddies, the QGPV flux is proportional to the wave activity tendency, as can be seen from equations (\ref{E:wave_activity}) and (\ref{E:gen_taylor}). In addition, in the WKBJ approximation, the EP flux is transporting wave activity at the group velocity. The applicability of the small-amplitude and WKBJ approximations in the troposphere is questionable (e.g., \citealp{potter2013}), Nevertheless, the QGPV flux can be interpreted as the tendency of a more general wave activity, defined for nonlinear eddies of any amplitude \citep{nakamura2010}. 
\begin{figure*}[ht!]
 \centering\includegraphics[width=140mm]{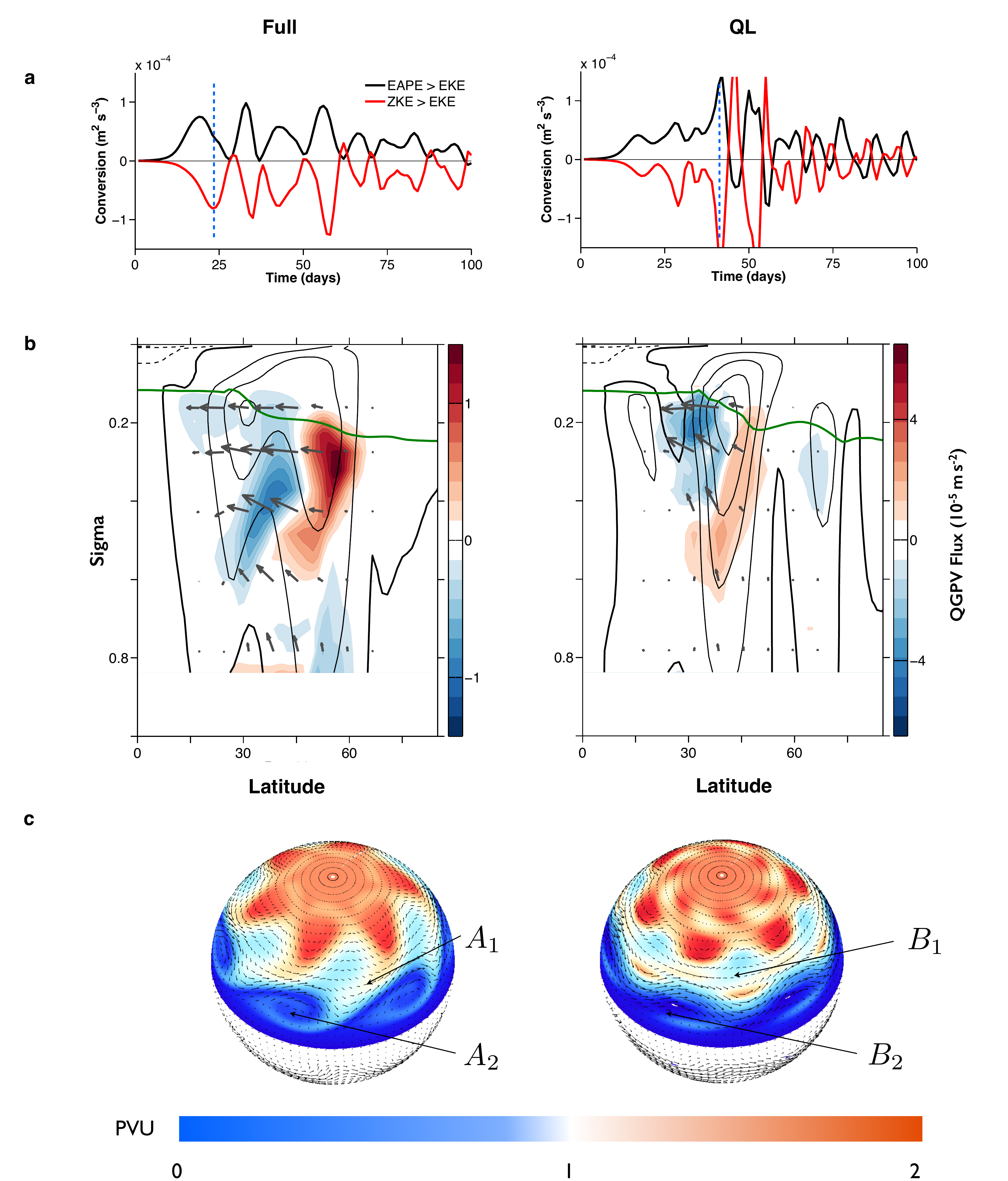}\\
 \caption{EP fluxes, QGPV flux, and PV when barotropic conversion of EKE to ZKE is maximal. Left column: full model; right column: QL model. (a) A reproduction of Fig.~\ref{fig:conv} with vertical dashed blue lines added to indicate when the EP flux and the QGPV flux in (b) are computed. (b) EP flux (grey arrows) and QGPV flux (colors). The EP flux and the QGPV flux are averaged over one day, between days 22 and 23 for the full model and between days 28 and 29 for the QL model. As in previous figures, solid contours are for eastward winds and dashed contours for westward winds, with $10 \unit{m\,s^{-1}}$ increments and the bold line indicating the zero-contour. The green line marks the tropopause. (c) Corresponding Rossby-Ertel PV maps on the 350K isentrope. The arrows are for the isentropic-density weighted winds.}\label{fig:LC_mid}
\end{figure*}
\begin{figure*}[ht!]
 \centering\includegraphics[width=140mm]{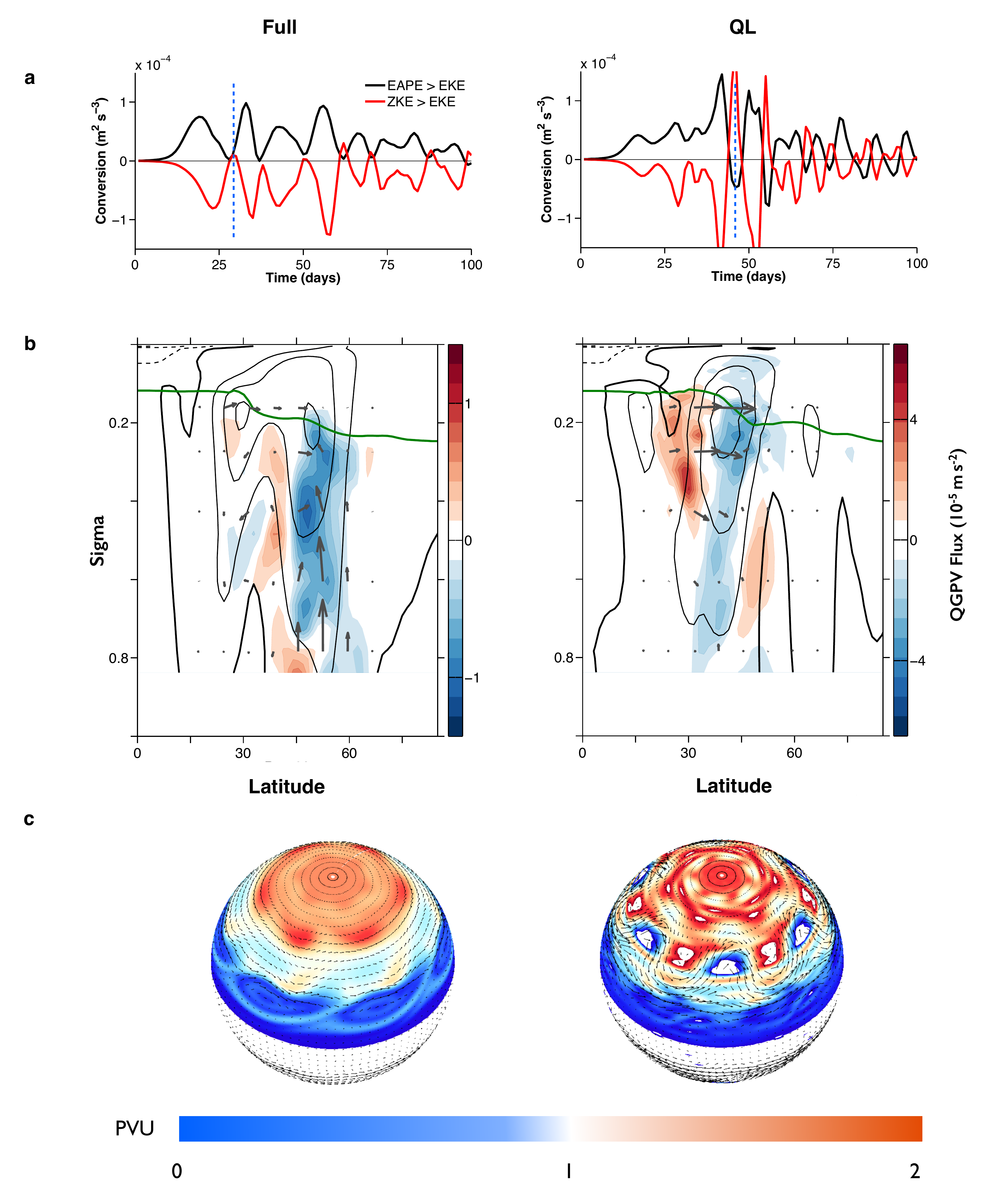}\\
 \caption{As in Fig.~\ref{fig:LC_mid} but at the end of a lifecyle, when the barotropic conversion of EKE to ZKE is minimal.}\label{fig:LC_end}
\end{figure*}

When the barotropic conversion of EKE to ZKE is maximal, the QL approximation captures the EP flux and wave activity tendencies fairly well (Fig.~\ref{fig:LC_mid}b). The QGPV flux in the middle to upper troposphere is negative, indicating the surf zone where eddies are absorbed. However, at the end of each lifecyle, when both barotropic and baroclinic conversions are weak, there is only weak re-emission of wave activity from the surf zone in the full model (Fig.~\ref{fig:LC_end}b, left panel), while there is very strong re-emission in the QL model, for example, at day 45 (Fig.~\ref{fig:LC_end}b, right panel). Wave activity radiation from the surf zone for QL dynamics results in barotropic re-growth of eddies (positive $C_K$ at day 45, Fig.~\ref{fig:LC_mid}a, right panel). Associated with positive barotropic conversion is baroclinic decay (negative $C_E$ at day 45, Fig.~\ref{fig:LC_end}a, right panel), consistent with an equatorward eddy flux of potential temperature and a downward EP flux (Fig.~\ref{fig:LC_end}b, right panel), unlike in the full model. 

The reason why wave activity on average is not absorbed in the QL upper troposphere (section~\ref{sec:noe}\ref{sec:mmc}) can now be better understood. Low-level baroclinic growth and vertical propagation of wave activity toward the tropopause are well captured. But after wave activity is absorbed in the upper-tropospheric surf zone, it is re-emitted. EP fluxes radiating from the tropical upper troposphere are particularly clear (Fig.~\ref{fig:LC_end}b, left panel). Similar absorption followed by re-emission also occurs on the poleward flank of the jet (not shown), although absorption there is of larger amplitude than emission and explains why EMF is concentrated on this flank of the jet in a statistically stationary state (Fig.~\ref{fig:edmf_wind}b). Inefficient absorption of wave activity but similar baroclinicity compared with the full model explains why EKE is larger in the QL model than in the full model (section~\ref{sec:noe}\ref{sec:mmc}): wave activity slushes meridionally in the upper troposphere, amplifying EKE and possibly leading to resonances within an upper-tropospheric wave guide.

The analysis of lifecycles confirms that lower-tropospheric dynamics are fairly well captured by the QL approximation and that the upper troposphere is more nonlinear than the lower troposphere (section~\ref{sec:noe}\ref{sec:mmc}). Bursts of baroclinic growth in the lower troposphere, consisting of alternate growth and decay of EP flux, are captured. This contrasts to some extent with previous studies of baroclinic lifecycles, in which low-level nonlinear saturation was invoked to explain why baroclinic conversion saturates \citep{simmons1978,heldhoskins1985,thorncroft1991}, as an essential step in a "saturation-propagation-staturation" paradigm for baroclinic wave lifecycles. Our lifecycle study suggests that QL mechanisms, like depletion of MAPE through baroclinic instability, might play a significant role in lower-tropospheric baroclinic growth saturation. However, strongly nonlinear mechanisms appear essential in the upper troposphere. It remains to discuss why upper-level eddy absorption is not fully captured in the QL model. 
 
\paragraph*{Potential vorticity rearrangement in the surf zone}\label{sec:critical_layer}

For adiabatic inviscid motion, potential vorticity is conserved on isentropes. Analyzing potential vorticity fields on isentropes can therefore be used to diagnose wave--mean flow interactions \citep{Hoskins1985}. Here we use the potential vorticity for the primitive-equations system (Rossby-Ertel, PV hereafter), rather than the QGPV. We evaluate it on the 350-K isentrope, which lies in the upper troposphere at low latitudes and in the lower stratosphere in polar regions (Fig.~\ref{fig:stream_wind}).

In a barotropic framework, Rossby waves can be absorbed quasi-linearly or nonlinearly \citep{held1987}. QL absorption occurs through the Orr mechanism, when a mean flow shears eddies in the same direction in which they are tilted and thereby transfers their energy to the mean flow \citep{farrell1987,lindzen1988}. As the eddies are sheared, the distance between vorticity extrema decreases, and vorticity anomalies become well approximated by the ratio of velocity anomalies to typical distances between neighboring extrema. Since this distance shrinks in the shearing process, velocity anomalies, and with them EKE, have to decreases because of vorticity conservation, leading to EKE transfer to the mean flow. The non-conservation of EKE is a direct consequence of the linearized dynamical operator acting on eddy fields being non-normal in the energy norm \citep{farrell1987,delsole2004}.

Eddy absorption, and the subsequent effect on the mean flow, is essentially a potential-vorticity mixing problem (see \citealp{dritschel2008} for a review), as expressed in the Taylor identity (\ref{E:gen_taylor}). For linear decay, PV mixing is performed by the shearing in the zonal direction of meridionally propagating eddies. However, PV mixing can be nonlinear, and in general is so in planetary atmospheres. Analytical theories were developed to understand absolute vorticity (or PV) rearrangement in nonlinear Rossby-wave breaking, notably the Stewartson-Warn-Warn (SWW) solution for barotropic inviscid critical layers of small-amplitude waves in a constant-shear mean flow \citep{stewartson1977,warn1978,killworth1985}. This theory predicts the formation of Kelvin cat's eye structures consisting of closed streamlines between fixed points near the critical lines, with a zonal wave number corresponding to the breaking wave. These structures are advecting PV anticyclonically at leading order as a passive scalar, forming small-scale PV filaments that are rolling up around the center of the cat's eye. Ultimately, this leads to vorticity mixing, which is here mediated by structures that are not zonally symmetric, in contrast to the linear decay. 

PV maps during maximal barotropic EKE-ZKE conversion are shown in Fig.~\ref{fig:LC_mid}c. They clearly show that wave activity absorption is nonlinear in the full model\footnote{The increasing amplitude of PV extrema between Figs. \ref{fig:LC_mid} and \ref{fig:LC_end} is worth noting and indicates that the numerical scheme does not conserve PV. The non-conservation is particularly striking in the QL approximation, probably because PV is not transferred as efficiently to small scales where it can be dissipated by hyperviscosity.}. The point $A_1$ marks a thinning PV filament, as described for LC1 wave breaking in \cite{thorncroft1991}. The region near $A_2$ resembles the Kelvin cat's eye of a critical layer, as predicted by SWW theory. Both phenomena involve eddy-eddy interactions and enstrophy cascading toward small scales. The filament $A_1$ constitutes the eastern flank of a structure also reminiscent of a cat's eye. Roll-up of filaments is not visible on the 350-K isentrope we are showing, but it is visible at lower levels. Corresponding structures arise in the QL model (near $B_{1}$ and $B_{2}$ on Fig.~\ref{fig:LC_mid}c). Barotropic linear theory of critical layers for small-amplitude waves indeed predicts the development of cat's eyes \citep{Dickinson1970}, before linear theory breaks down. QL dynamics cannot capture the subsequent thinning and roll-up of PV filaments, which are essential for vorticity mixing and the absorption of wave activity in critical layers. However, QL dynamics does capture the formation of cat's eyes \citep{Haynes1987}. The baroclinic structures we observe here are more complex, but the simplified barotropic small-amplitude framework elucidates why they develop. 

In contrast to the full model, in the QL model, dipoles of positive and negative PV anomalies near $B_{1}$ and $B_{2}$ in Fig.~\ref{fig:LC_end} persist because eddy-eddy interactions are not available to excite zonal wavenumbers larger than that of the breaking wave to allow filamentation. \cite{Haynes1987} describe a similar phenomenon for the QL SWW solution. The rotation of PV anomalies around the center of the cat's eye leads to alternate phases of absorption and overreflection of the same amplitude (instead of decreasing amplitude implied by filamentation in the fully nonlinear case). Re-emission of wave activity in our experiments is the baroclinic large-amplitude equivalent of the overreflection phase. When the positive vorticity anomaly is on the eastern side of the cat's eye, the tilt is southwest to northeast such that EMF is poleward. When the vorticity anomaly is advected to the west of the cat's eye, the tilt becomes southeast to northwest because it joins with the positive anomaly northwest of the cat's eye. Hence, EMF is equatorward. This is the essence of the overreflection phase. Because the phase of absorption, in which the correlation between $u'$ and $v'$ is positive, is followed by a phase of overreflection, in which the correlation is negative, the correlation is on average close to zero, as observed in the statistically stationary circulation (section~\ref{sec:noe}\ref{sec:edmf}).

It is important to stress that nonlinear structures mediating PV rearrangement (cat's eyes) have a large meridional extent, spanning much of the baroclinic zone in both the full and the QL simulations. The small-amplitude calculations of \cite{Dickinson1970} and \cite{Haynes1987} are consistent with our finite-amplitude computation: similar cat's eye structures arise in both the nonlinear and the QL simulations. The fundamental difference between the two cases lies in the details of vorticity dynamics inside the cat's eyes.

PV maps at the end of a lifecycle are shown in Fig.~\ref{fig:LC_end}c. Weak meridional gradients and small-amplitude zonal structures in the fully nonlinear simulation are a consequence of eddies having been absorbed. In the QL model, large-amplitude waves and strong PV gradients indicate wave resonance, resulting from wave activity re-emission. The absence of eddy-eddy interactions prevents PV filamentation and transfer of enstrophy and of wave activity toward small scales, where they can be dissipated. However, the shortcomings of the QL model are already manifest within the framework of conservative dynamics. For barotropic SWW critical layers, this is shown in \cite{Haynes1987}. The exact role of diffusion in wave absorption in a more complex system is not clear. 

EMF is concentrated on the poleward flank of the jet in the QL simulation (Fig.~\ref{fig:edmf_wind}). More work is required to understand this fact. We can conjecture that QL absorption through the Orr mechanism is more efficient on the poleward flank of the jet. Ineed, PV maps do not show the formation of cat's eye on this flank of the jet but only the shearing of eddies by the mean flow. 

\subsection{The role of barotropic triads}\label{sec:bt_triads}

\begin{figure}[t]
  \centering\includegraphics[width=19pc]{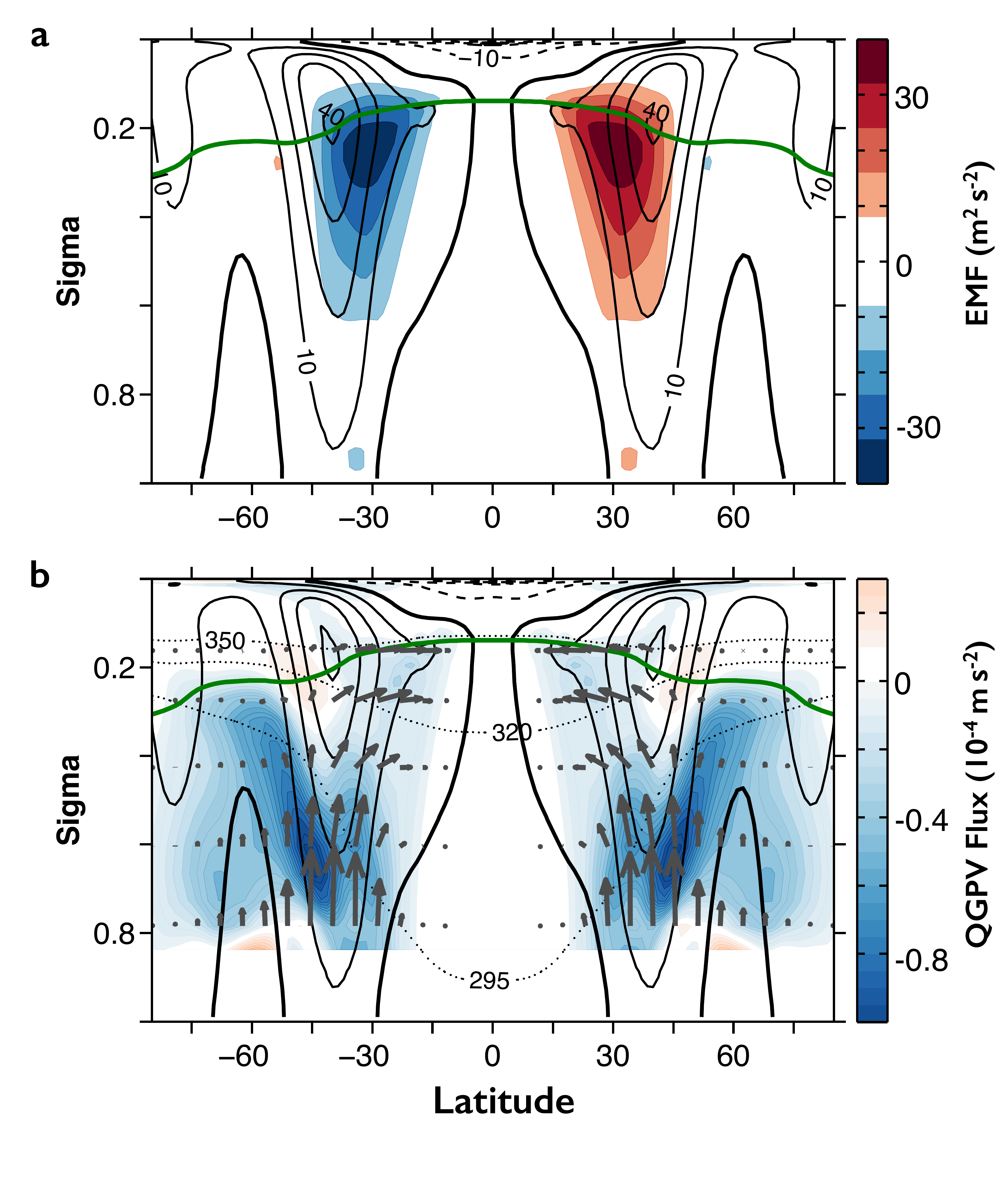}\\
  \caption{Simplified GCM with wave-mean flow and barotropic eddy-eddy interactions, as described in section~\ref{sec:noe}\ref{sec:bt_triads}. (a) Zonal wind and EMF (as in Fig.~\ref{fig:edmf_wind}). (b) EP  vector and QGPV flux (as in Fig.~\ref{fig:vq_EP_wind}).}\label{fig:diag_noe_bt}
\end{figure}

The lifecycle calculations show that wave decay is primarily nonlinear for Earth-like parameters. To determine what components of atmospheric turbulence are crucial, we have also considered a simplified GCM in which only barotropic triads are retained, while baroclinic-baroclinic triads and baroclinic-barotropic triads are neglected. Turbulence, here taken to mean transfer of inviscid quadratic invariants among scales, is contained in the barotropic mode only.  Defining a vertical average with brackets $[\cdot]$ and a zonal average with an overline, the time tendency of a scalar $a=\bar{a}+a'$ due to the meridional advection by the velocity $v=\bar{v}+v'$ is integrated as follows (cf. Eq. \ref{E:advection_noe}):
\begin{align}\label{E:advection_noe_bc}
	\frac{\partial a}{\partial t} =
    -\bar{v}\frac{\partial \bar{a}}{\partial y}-
	\bar{v}\frac{\partial a'}{\partial y}-
	v'\frac{\partial \bar{a}}{\partial y}&-
	\overline{v'\frac{\partial a'}{\partial y}}  \notag \\
    & -\left( [v']\frac{\partial [a']}{\partial y}-\overline{[v']\frac{\partial [a']}{\partial y}} \right) \mbox{.}
\end{align}
Restoring the barotropic triads gives an EMF structure that is much closer to the full model. The upper-tropospheric enhancement, with realistic amplitudes, is captured, as can be seen in the upper panel of Fig.~\ref{fig:diag_noe_bt}. This is consistent with EP vectors extending toward the subtropical upper troposphere and showing eddy absorption at the equatorward flank of the jet (Fig.~\ref{fig:diag_noe_bt}b). Nevertheless, eddies are still compressed in the meridional direction, and secondary eddy-driven jets are developing, though at significantly higher latitudes than in the QL model. Eddy absorption on the poleward flank of the jet is reduced compared with the QL model but is still more significant than in the full model (not shown). Lifecycle experiments confirm that nonlinear saturation does occur on the equatorward flank of the jet. Nevertheless, absorption on the equatorward flank is less efficient than in the full model. Significant wave activity is still re-emitted and then absorbed on the poleward flank of the jet. As a consquence, EMF divergence is stronger than in the full model on the poleward flank of the jet. This is likely linked to the compression in the meridional direction of the circulation. Nonetheless, it is clear that allowing the nonlinear interaction of barotropic triads suffices to obtain a much more realistic baroclinic EMF structure. 

It is not clear how the baroclinic waves interact with critical layers and why barotropic interactions suffice to give the baroclinic EMF structure. The baroclinic-baroclinic interactions are likely not important because vertical wavenumber one dominates the vertical structure of the atmosphere: baroclinic-baroclinic interactions only can affect wavenumbers larger than two. However baroclinic-barotropic triads are a priori important. Neglecting baroclinic-baroclinic triads and barotropic-barotropic triads, while retaining baroclinic-barotropic triads, also captures upper-level enhancement. Thus, although dissipation of wave activity is not simply related to the barotropic (vertically avergaed) flow, retaining barotropic interactions (barotropic-barotropic triads or baroclinic-barotropic triads) suffices to capture some eddy absorption. However, it does not capture the nonlinear dynamics fully.
%


\section{Why eddy momentum fluxes are concentrated in the upper troposphere}\label{sec:mech}
 
\subsection{Summary and discussion}

In section \ref{sec:noe}, we have shown that EMF is not enhanced in the upper troposphere in the QL model with Earth-like parameters because wave activity is not absorbed on the equatorward flank of the jet stream. Baroclinic eddies can be absorbed through two fundamentally different mechanisms: nonlinear saturation and the QL Orr mechanism. Lifecycle experiments show that for Earth-like parameters, and for rapidly growing baroclinic eddies, nonlinear saturation is more relevant. While the Orr mechanism can be captured by QL dynamics, nonlinear saturation cannot. Essentially, eddy-eddy interactions allow wave activity absorption through PV rearrangement in the surf zone and wave activity dissipation through enstrophy transfer to small scales. Without eddy-eddy interactions, wave activity is primarily re-emitted from the surf zone, leading to excessive EKE in the upper troposphere. 

Yet it is possible to construct circulations in which eddy amplitudes or the geometry of the mean flow favor QL wave decay over nonlinear saturation, for example, by decreasing the pole-to-equator contrast of the radiative forcing, by decreasing surface friction, or by making the planet rotate faster (not shown). Under such circumstances, the QL approximation performs better and captures upper-level EMF enhancement more accurately. In some cases, then, fundamental aspects of the tropospheric EMF vertical structure can be understood in terms of QL dynamics, without having to consider nonlinear mechanisms for wave absorption. Wave activity is generated linearly in the lower troposphere and propagates vertically. Then, the tropopause acts as a turning surface for Rossby waves, trapping them in the troposphere and guiding them to propagate meridionally. This explains why no substantial EMF extends above the tropopause. In our experiment in which the poles were heated (section~\ref{sec:surf_fric}), waves are propagating downward, and the solid surface plays a similar role, resulting in EMF concentration near the surface.

\subsection{Depth of baroclinic eddies}

There is a vertical level below which EMF is weak (Fig.~\ref{fig:edmf_wind}). The EP flux is primarily vertical in the lower troposphere because baroclinic growth dominates the dynamics. One question arises: above which altitude does the EP flux acquire a substantial meridional component? That is, at which altitude does EMF become significant? We suggest this level to be controlled by the typical vertical extent of baroclinic eddies \citep{held1978,schneider2006}.

To verify this hypothesis, we compare simulations in which the midlatitude tropopause height is set by convection with simulations in which it is set by the typical depth of baroclinic eddies \citep{schneider2004b,schneider2006}. This is achieved by changing the lapse rate $\gamma \,g/c_p$ to which the convective parametrization is relaxing temperature profiles in our idealized GCM (section \ref{sec:GCM}). As the convective lapse rate is reduced ($\gamma$ gets smaller), the tropopause rises, and convection becomes increasingly important for the extratropical thermal stratification \citep{schneider2006,schneider2008b,ogorman2011},  setting the height of the tropopause for $\gamma \lesssim 0.6$. To resolve the upper troposphere and lower stratosphere well as the tropopause height increases, we perform these simulations with 60 vertical $\sigma$-levels, instead of 30 levels in the previous simulations. 

\begin{figure*}[ht!]
  \centering\includegraphics[width=\textwidth]{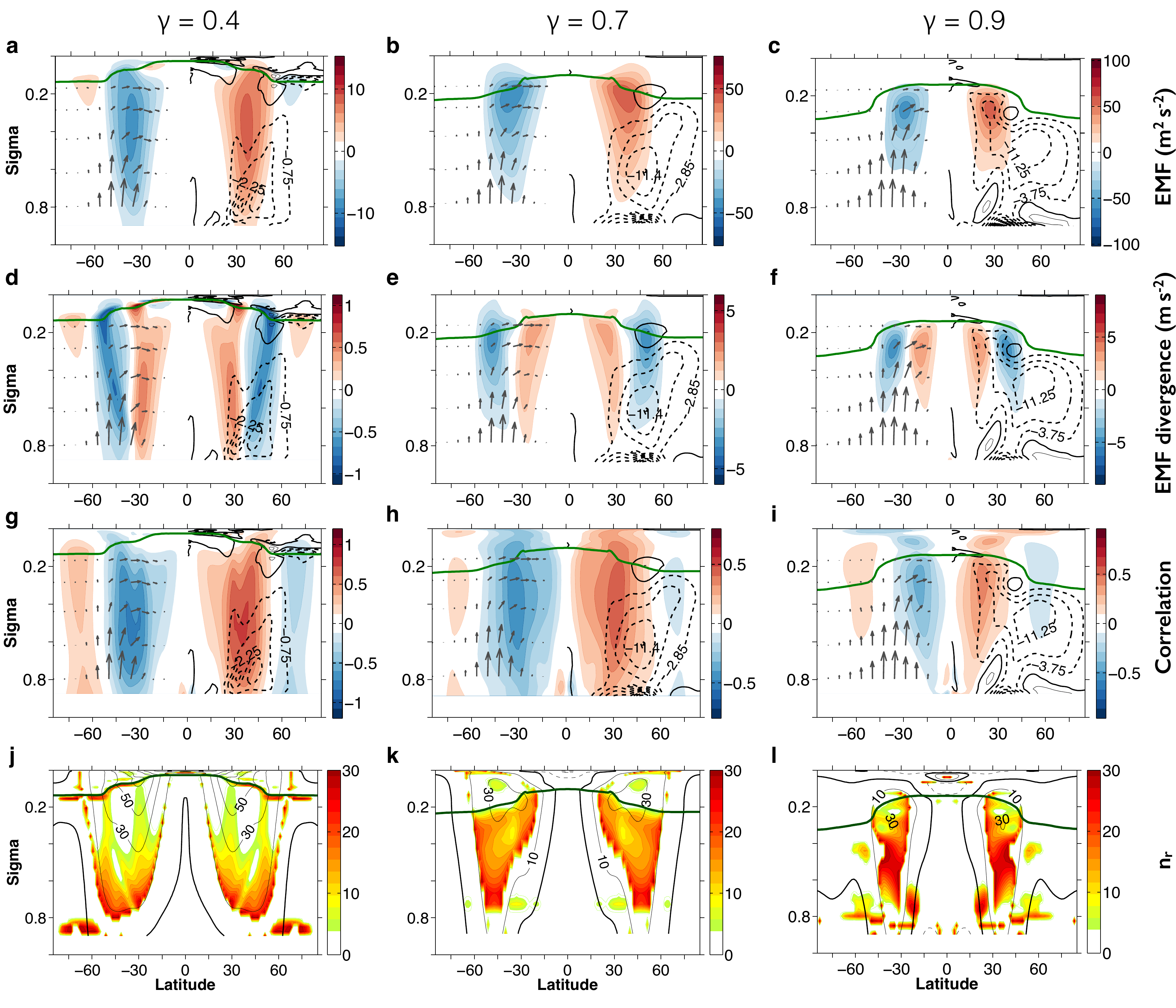}\\
  \caption{(a-c) EMF (colors), QGPV flux (dashed contours for negative values and solid contours for positive values, in $10^{-5} \unit{m\,s^{-2}}$, in the northern hemisphere only), and the EP flux (grey arrows for midlatitudes in the southern hemisphere only). (d-f) EMF divergence (colors), QGPV flux (contours, as in (a-c)) and EP flux (arrows, as in (a-c)). (g-i) Correlation $\overline{u'\,v'}/(\overline{u'^2}\,\overline{v'^2})^\frac{1}{2}$, QGPV flux (contours, as in (a-c)) and EP flux (arrows, as in (a-c)). Note the change of scales for different values of $\gamma$. (j-l) Rossby-wave refractive indices $n_r$ (colors) and zonal mean wind (contours, in $\unit{m\,s^{-1}}$). In all figures, the thick green line marks the tropopause (a 2-\unit{K\,km^{-1}} lapse rate contour). Columns from left to right: $\gamma=0.4$, $\gamma=0.7$ and $\gamma=0.9$. }\label{fig:summary_gamma}
\end{figure*}

The EMF for simulations with $0.4 \leq \gamma \leq 0.9$ is shown in Fig.~\ref{fig:summary_gamma}a-c, together with the EP vectors and the QGPV flux in dashed contours to indicate the depth of baroclinic eddies. The convective lapse rate parameter $\gamma$ has an important effect on the EMF structure: as $\gamma$ is increased and baroclinic eddies become more important in controlling the extratropical thermal stratification, EMF become more peaked in the upper troposphere. 

For small convective lapse rates (e.g., $\gamma=0.4$), the EMF is strong over a large vertical extent and is maximal well below the tropopause (Fig.~\ref{fig:summary_gamma}a). The EMF divergence (Fig.~\ref{fig:summary_gamma}b) exhibits a particularly complex structure, in comparison to larger $\gamma$, because it has two distinct local maxima: in the mid-troposphere ($\sigma \approx 0.5$) and near the tropopause ($\sigma \approx 0.1$).

EMF convergence also exhibits two corresponding maxima, but the one in the mid-troposphere is weak. To elucidate how this structure arises, we show in Fig.~\ref{fig:summary_gamma}g the correlation coefficient between meridional and zonal velocity anomalies $\overline{u'v'}/[\overline{u'^2}\,\overline{v'^2}]^{1/2}$, where the overbar here stands for a time and zonal average. It appears that the absolute value of the correlation coefficient only has one clear mid-latitude maximum, near the level of the mid-tropospheric EMF convergence/divergence extrema. Large absolute values of the correlation coefficient indicate large EMF relative to eddy amplitude (EKE), and thus significant eddy absorption. Hence, comparing Figs.~\ref{fig:summary_gamma}d and \ref{fig:summary_gamma}a suggests that the mid-tropospheric EMF convergence/divergence extrema are caused by the absorption of baroclinic eddies originating from lower levels, while the mechanisms responsible for the tropopause extrema are distinct. In fact, the EMF convergence and divergence near the tropopause correspond to northward and southward fluxes of PV (the potential temperature flux contribution to the QGPV flux there is weak, unlike in lower tropospheric layers). The northward  PV flux seems upgradient, and such fluxes have recently been argued to arise from nonlinear wave breaking \citep{birner2013}. The circulation with $\gamma = 0.4$ challenges this explanation because wave breaking regions appear well below the tropopause. Other processes may be responsible for the northward near-tropopause QGPV flux, such as local barotropic instability involving shallow modes.

As the convective lapse rate ($\gamma$) increases, the altitude where the absolute value of the correlation coefficient between $u'$ and $v'$ in midlatitudes is maximal also increases, occurring at $\sigma = 0.55$, $0.4$, and $0.3$ for $\gamma=0.4$, $0.7$, and $0.9$ (Fig.~\ref{fig:summary_gamma}g-i). This closely follows the deepening of baroclinic eddies, as shown by the tropospheric QGPV flux (e.g., Fig.~\ref{fig:summary_gamma}g-i). The level up to which the EP flux penetrates and the level of maximum correlation roughly coincide. The mid-tropospheric maximum of EMF divergence at the vertical level of maximum correlation disappears as $\gamma$ is increased from $0.4$ to $0.7$, leaving only the near-tropopause extrema. Apparently, the gap between the tropopause and the penetration depth of baroclinic eddies shrinks as $\gamma$ increases, both being equal for sufficiently large convective lapse rates ($\gamma \gtrsim 0.7$, see Fig.~\ref{fig:summary_gamma}a-c). But only for $\gamma=0.9$ does the vertical structure of the correlation coefficient resembles that of the EMF. The dynamics that account for the near-tropopause EMF divergence/convergence extrema at $\gamma = 0.4$, and contribute to it at larger $\gamma$, do not leave a noticable signature in the correlation coefficient; they appear to be distinct from the lower-tropospheric baroclinic-eddy dynamics. A simulation with $\gamma=0.4$, in which the vertical resolution is halved, accurately captures the mid-tropospheric EMF divergence/convergence extrema but does not exhibit the near-tropopause maxima. This also points to  distinct, shallow mechanisms being responsible for the near-tropopause EMF divergence/convergence extrema for $\gamma \lesssim 0.7$.

To diagnose where the mean flow allows Rossby-wave propagation, and thus how deeply baroclinic wave can penetrate, we use refractive indices as defined in \cite{seager2003}:
\begin{equation}\label{E:ref_index}
	n_r^2 = \frac{R^2\bar{q}_{y}}{\overline{u}-\Re(\sigma)R\cos\phi/{k_i}} - \frac{k_i^2}{\cos\,\phi} + R^2 F(N)
\end{equation}
where, in $z$-coordinates \citep{harnik2001}:
\begin{equation}\label{E:ref_index}
	F(N)= f^2 \frac{e^{z/2h}}{N} \frac{\partial}{\partial z} \left[\frac{e^{-z/h}}{N^2}\frac{\partial}{\partial z}(e^{z/2h}N)\right]\mbox{.}
\end{equation}
We denote with $N$ and $h$ the Brunt-V\"ais\"al\"a frequency and the pressure scale height. The refractive index is obtained by assuming a normal mode solution $\Psi(\phi,z)e^{i(k_i\lambda-\sigma t)}$ to the linearized QGPV equation around a zonal mean flow of QGPV $\bar{q}$ ($\lambda$ and $k_i$ are the longitude and zonal wavenumber). The resulting wave equation yields \citep{harnik2001}:
\begin{equation}\label{E:wave_eq}
	\nabla^2\Psi + n_r^2 \Psi = \text{forcing and disssipation}\mbox{.}
\end{equation}
The vertical coordinate in the two-dimensional laplacian operator of Eq. (\ref{E:wave_eq}) is rescaled by the Prandtl ratio $f/N$. Positive values of $n_r^2$ indicate the possibility for wave propagation, whereas negative index preclude it. We compute the refractive indices for the fastest growing normal mode on the zonal- and time-mean circulation. 

Rossby-wave refractive indices are consistent with baroclinic eddies not reaching the tropopause for $\gamma=0.4$: the upper-tropospheric refractive index is negative near the jets with turning surfaces $n^2_{r}=0$ in the mid-troposphere (Fig.~\ref{fig:summary_gamma}j). For $\gamma=0.4$, baroclinic eddies are shallower than the tropopause; they do not propagate sufficiently high vertically to reach it. This is consistent with the EMF maxima matching the depth of baroclinic eddies, and it further indicates that EMF near the tropopause is due to a shallow local mechanism 

To summarize, the concentration of EMF in the upper troposphere emerges for seemingly two distinct reasons. First, baroclinic eddies appear to be absorbed at a height that is determined by their typical depth. When this depth corresponds to the tropopause height ($\gamma \gtrsim 0.7)$, EMF is enhanced in the upper troposphere.  Second, near-tropopause dynamics are responsible for some meridional momentum flux (and even stronger EKE relatively to lower levels baroclinic activity); the dynamical processes responsible for this are unclear but appear to be local to the tropopause. 


\section{Conclusions}

Using an idealized dry GCM, we have investigated how the vertical structure of EMF is controlled and how its concentration in the upper troposphere arises. In a simulation in which the poles were heated relative to the equator, we obtained an upside-down version of Earth's tropospheric circulation with EMF and EKE enhanced close to the surface. This shows that surface friction is not responsible for Earth's weak EMF near the surface, as had been suggested  \citep{held2000,vallis2006}. Nonlinear lifecycle experiments are consistent with this conclusion in that they exhibit enhanced EMF in the upper troposphere, although surface friction in them is disabled.

The upper atmosphere favors linear Rossby-wave propagation more than the lower troposphere. It has been suggested that this explains the EMF asymmetry between the upper and lower troposphere \citep{held2007}. To test this hypothesis, we compared a fully nonlinear model to a QL model, in which interactions between eddies and the mean flow are retained while nonlinear eddy-eddy interactions are suppressed. We have shown that the QL model, despite some success in capturing important aspects or planetary large-scale dynamics \citep{ogorman2007}, does not reproduce the vertical EMF structure and its concentration in the upper troposphere. The reason is that eddy absorption in Earth-like parameter regimes is strongly nonlinear: nonlinear eddy-eddy interactions in the surf zone in the upper troposphere are important for the absorption of wave activity. Wave activity is re-emitted from the surf zone when eddy-eddy interactions are suppressed, as we saw in the QL baroclinic lifecycle experiments. This results in excessive eddy kinetic energy in the upper troposphere. That is, although the upper troposphere appears relatively linear as far as wave propagation characteristics are concerned \citep{randel1991}, it is more nonlinear than the lower troposphere with regards to eddy absorption. The QL model captures lower-tropospheric dynamics (e.g., baroclinic growth and its saturation) more faithfully than upper-tropospheric dynamics (e.g., the nonlinear surf zone). The QL model does not capture the essence of the upper-tropospheric dynamics because the surf zone in which nonlinear eddy-eddy interaction are important is not small in the meridional direction (as it is in some idealized flows like SWW); rather, it spans the meridional width of the baroclinic zone. An important consequence is that there is no clear separation between the latitude of generation of baroclinic eddies and the surf zone. EMF in this case occurs when dissipation occurs in the wings of eddies.

To understand the vertical EMF structure, it is fruitful to think in terms of generation, propagation, and absorption of wave activity. Wave activity is generated in the lower troposphere, propagates upward, then turns meridionally and is absorbed in the upper troposphere \citep{simmons1978,simmons1980,thorncroft1991}, preferentially at a level that scales with the typical depth of baroclinic eddies. Wave activity propagation above the tropopause is inhibited because of the strong stabilization of the stratification above it, leading to the tropopause acting as  a turning surface or wave guide \citep{thorncroft1991}. As a consequence, on Earth, the EMF structure is peaked in the upper troposphere because the tropopause height and the typical depth of baroclinic eddies coincide \citep{held1982, schneider2004b,schneider2006,ogorman2011}. We have shown that when the tropopause is set by convection and baroclinic eddies do not reach the tropopause, EMF has a more complex structure with a maximum well below the tropopause that corresponds to the typical depth of baroclinic eddies. Additionally, there can be near-tropopause maxima of EMF divergence/convergence generated by distinct dynamics, whose origin is not entirely clear but which also play some role in upper-tropospheric EMF enhancement when the tropopause height is set by baroclinic eddies. Similarly but with reversed sign, in the simulation in which the poles were heated more strongly than the equator, wave activity is generated in the mid-latitude troposphere, propagates downward, then turns meridionally and is absorbed near the surface. In this case, the surface plays the role of a turning surface, with EMF reaching its maximum in the lower troposphere.  

A more complete understanding of the EMF structure in baroclinic atmospheres would require an explanation of why wave-activity absorption occurs preferentially at a height scaling with the typical depth of baroclinic eddies. Also, near-tropopause dynamics should be investigated further because they seem to be responsible for significant EMF and so affect the large-scale circulation.

The results obtained with the QL GCM have implications for the development of statistical closures for large-scale atmospheric dynamics. What favors quasi-linear or nonlinear decay, even in a simple barotropic framework, and what determines their relative importance when both can occur, is not fully understood. Unraveling this would be an essential step to develop such closures. Successes of second-order closures based on QL dynamics are likely explained by quasi-linear absorption being favored because of the structure of the flow \citep{farrell1996,farrell1996b,bouchet2013} or because of model tuning \citep{whitaker1998,zhang1999,delsole2001}. The central role of eddy-eddy interactions in planetary macroturbulence does not contradict weakly nonlinear descriptions and theories, for example of the thermal structure \citep{randel1991,held2000,schneider2006,schneider2008}, in the sense that nonlinear eddy-eddy interactions only are essential for eddy absorption. Somewhat ironically, the atmosphere looks more linear in the presence of nonlinear eddy-eddy interactions than in the QL approximation: eddies are of weaker amplitude with respect to the mean flow in the full model than in the QL model because their absorption is inhibited in the QL model. 

%
\acknowledgments
This work was supported by the US National Science Foundation grants CCF-$1048575$ and CCF-$1048701$. We thank Brad Marston for useful discussions about quasi-linear approaches and for suggesting investigating the role of barotropic triads (section~\ref{sec:noe}\ref{sec:bt_triads}). We also thank Freddy Bouchet and Cesare Nardini for useful discussions about the Orr mechanism at the Kavli Institute for Theoretical Physics (KITP) summer 2014 program on wave--flow interaction in geophysics, climate, astrophysics, and plasmas. 

%






%


 \bibliographystyle{ametsoc2014}
 \bibliography{references}

%

%

\end{document}